\documentclass[a4paper,onecolumn,prd,aps,accepted=2020-06-28]{quantumarticle}
\pdfoutput=1

\usepackage{fancyhdr}
\usepackage[margin=2.2cm]{geometry}
\usepackage{parskip}
\usepackage{tensor}
\usepackage{amsmath}
\usepackage{amssymb}
\usepackage{amsthm}
\usepackage{bbm}
\usepackage{physics}
\usepackage{slashed}
\usepackage[numbers,sort&compress]{natbib}
\usepackage{graphicx}
\usepackage{url}

\begin{document}
\title{Classical Simulations of Quantum Field Theory
in Curved Spacetime I: Fermionic Hawking-Hartle Vacua from a Staggered Lattice Scheme}
\author{Adam G. M. Lewis}
\affiliation{Perimeter Institute for Theoretical Physics, 31 Caroline St. N., Waterloo, Ontario, Canada, N2L 2Y5}
\email{alewis@perimeterinstitute.ca}
\author{Guifr\'{e} Vidal}
\affiliation{Perimeter Institute for Theoretical Physics, 31 Caroline St. N., Waterloo, Ontario, Canada, N2L 2Y5}
\affiliation{X, The Moonshot Factory, Mountain View, CA 94043}
\maketitle
\begin{abstract}
We numerically compute renormalized expectation values of quadratic operators in a quantum field theory (QFT) of free Dirac fermions in curved two-dimensional (Lorentzian) spacetime. 
First, we use a staggered-fermion discretization to generate a sequence of lattice theories yielding the desired QFT in the continuum limit. 
Numerically-computed lattice correlators are then used to approximate, through extrapolation, those in the continuum. 
Finally, we use so-called point-splitting regularization and Hadamard renormalization to remove divergences, and thus obtain finite, renormalized  expectation values of quadratic operators in the continuum.
As illustrative applications, we show how to recover the Unruh effect in flat spacetime and how to compute renormalized expectation values in the Hawking-Hartle vacuum of a Schwarzschild black hole and in the Bunch-Davies vacuum of an expanding universe described by de Sitter spacetime.
Although here we address a non-interacting QFT using free fermion techniques, the framework described in this paper lays the groundwork for a series of subsequent studies involving simulation of interacting QFTs in curved spacetime by tensor network techniques. 
\end{abstract}

\section{Introduction}
While by now it is generally expected that black holes will radiate and evaporate \cite{HawkingRad1974, PhysRevD.14.870, PhysRevD.45.R1005} in the presence of a quantum field,
the details remain murky. The specific interplay between gravity and quantum fields during the early Universe's presumed inflationary \cite{OldInflation1, OldInflation2, NewInflation1, NewInflation2} phase
is similarly unclear. This is not (necessarily)
because we lack a quantum theory of gravity: one expects that at least some of these effects might already be acceptably described in some sort of \textit{semiclassical regime}. For example, one might consider the semiclassical field equations
\begin{equation}
\label{eq:Semiclassical}
G_{\mu \nu} = \frac{1}{8 \pi} \langle T^\mathrm{ren}_{\mu \nu} \rangle
\end{equation}
that relate the classical Einstein tensor $G_{\mu \nu}$  to the expectation value $\langle T^\mathrm{ren}_{\mu \nu} \rangle$ of a quantum field's suitably-renormalized stress-energy tensor $T_{\mu \nu}$. Given this regime indeed exists, the real problem is that \eqref{eq:Semiclassical} is hard to solve. On the left hand side, we have the fully non-linear Einstein equations, while on the right sits an out-of-equilibrium, perhaps strongly-interacting, quantum field theory in curved spacetime. Even taken separately, such problems are typically intractable.

But much further progress can be made with either \emph{numerically}. Since initial breakthroughs in 2005, \cite{pretorius2005numerical, pretorius2005evolution} fully nonlinear simulations of general relativity are now routine. The quantum field theory is a bit trickier, due to the inability of classical hardware to represent highly entangled wavefunctions. In two spacetime dimensions, however, a tensor network ansatz known as the \textit{matrix product state} \cite{Fannes_1992, White1992, White1993, OstlundRommer1995, OstlundRommer1997, VidalMPS1, VidalMPS2, Daley2004, White2004, PerezGarciaVerstraete2007,  Schollwock2005, Schollwock2011} can efficiently simulate low-energy states of local Hamiltonians, as well as sufficiently short real-time evolutions of them.

This is the first in a series of studies attacking \eqref{eq:Semiclassical} by combining numerical relativity with tensor network approaches. The first step is to write down the Hamiltonian evolving the quantum field forward in some time coordinate. This Hamiltonian is then mapped to the continuum limit of a sequence of lattice Hamiltonians. On each lattice, interesting states such as the ground state and other low energy states, as well as thermal states and their real-time evolutions, will be simulated with matrix product state techniques. From that sequence we extract $\langle T^\mathrm{ren}_{\mu \nu}\rangle$, the (renormalized) continuum limit of the analogous lattice expectation value. The expectation value $\langle T^\mathrm{ren}_{\mu \nu}\rangle$ can then serve as the source term in a numerical relativity simulation.

A necessary first step towards the use of tensor networks such as matrix product states is a careful translation of the continuum QFT problem into one on the lattice. This paper documents that undertaking: the mapping between expectation values in a QFT on a \emph{fixed} curved spacetime \cite{BirrellAndDavies, WaldQFT, ParkerAndToms} and the aforementioned renormalized sequence of lattice expectation values. 

We study Dirac fermions in a curved, two-dimensional Lorentzian spacetime. Dirac fermions in two spacetime dimensions are among the easiest fields to study on the lattice, because they map straightforwardly to an approximation by \textit{staggered} lattice fermions \cite{KogutSusskindLattice, Susskind_Lattice_1976, SusskindLattice1976} which in turn can mapped exactly to a quantum spin chain \cite{JordanWigner}.
The quantum spin chain can then be simulated by matrix product state techniques without further ado, as we will demonstrate in \cite{InProgress}. In the current derivations, however, in order to not further complicate an already rather involved discussion, we consider a \textit{non-interacting} Dirac field, which can be solved without resorting to matrix products states, using the free-fermion formalism. We note this part of the program is comparable to that proposed in \cite{yang2019simulate} and, more closely, \cite{GhentThesis}, which studied lattice Hamiltonians equivalent to the one we eventually derive. 

In general spacetimes it is not always clear what states ought to be of physical interest. In the literature one often proceeds by making mode expansions of field operators, whose forms are somehow determined by the problem at hand. These mode expansions might not necessarily relate straightforwardly to the Hamiltonian formulation we use. To make comparisons with results based on them, we consider a class of states called ``Hawking-Hartle vacua'' \cite{HawkingHartle, ISRAEL1976, GibbonsPerry, UnruhWeiss, LAFLAMME1989, KayWald1991, JacobsonNote}. They
are defined in spacetimes containing static observers, bounded by a so-called ``Killing horizon". In the Hawking-Hartle vacuum, that horizon radiates, but is at thermal equilibrium with its environment. It can be viewed as a thermal state of the Hamiltonian defined by static observers outside the horizon, at a certain ``Unruh'' temperature. We identify it (in the continuum limit) with thermal
states of the analogous lattice Hamiltonian, at the relevant metric's Unruh temperature.

In a quantum field theory, the expectation value $\langle T_{\mu \nu} \rangle$ of the stress-energy tensor $T_{\mu\nu}$ in the Hawking-Hartle vacuum is formally infinite even for free fields, and to extract a meaningful finite number it needs to be renormalized. This same problem occurs for any operator with quadratic terms in the fields. One possible solution, formulated directly in coordinate space and thus appropriate to curved spacetime, is Hadamard renormalization \cite{hadamardCauchy, ChristensenReg1978, 1984Ottewill, DEWITT1960, ChristensenVev1976, DecaniniFolacci, AMBRUS2015, lewis2019hadamard}. Here, the quadratic operator is replaced with the \textit{coincidence} (zero-separation) limit of some correlation function, with any divergent terms subtracted before the limit is taken. In fact, the lattice two point functions have the same leading order form as their continuum limits in a short-distance expansion. Thus, substituting them for the continuum correlation functions within the coincidence limit yields the same result. 

For illustrative purposes we present four applications in the continuum using lattice data. By order of complexity, they are: (i) computation of ground state expectation values as measured by inertial Minkowski observers; (ii) computation of thermal expectation values as measured by accelerated Minkowski observers (and thus reproduction of the Unruh effect \cite{PhysRevD.14.870}); (iii) computation of thermal expectation values as measured by static observers in the Hawking-Hartle vaccum of Schwarzschild black hole spacetime; and, finally, (iv) computation of thermal expectation values as measured by static observers in the Hawking-Hartle vacuum (often called the Bunch-Davies vacuum) of an expanding universe described by de Sitter spacetime  \cite{BunchDavies, AllendS, AIHPA_1968__9_2_109_0, Greene_2006, PhysRevD.100.085002, HawkingCosmo}.

This paper is divided into four main sections. The first is this introduction. The second develops the continuum problem to be numerically solved: it defines the free Dirac field in a curved two-dimensional spacetime, explicates its canonical quantization, and details how point-splitting regularization and Hadamard renormalization allow quadratic expectation values to be defined. It also introduces examples of spacetimes with a bifurcate Killing horizon and defines the Hawking-Hartle vacuum. 
The third section outlines our lattice approximation: it introduces the lattice and then reviews the staggered fermion technique to map between its Hamiltonian and that of the continuous field theory, as well as the \textit{free fermion} formalism to find correlation functions of non-interacting fermions on the lattice. 
Finally, the fourth section uses the above method to find Hadamard-renormalized expectation values in the four example cases (i)-(iv) previously listed.

\section{Continuum Problem}

To obtain $\langle T^\mathrm{ren}_{\mu \nu} \rangle$ in \eqref{eq:Semiclassical}, we must understand how to compute renormalized expectation values of \textit{quadratic operators}. The goal of this paper is to explain how to do so for so-called ``Hawking-Hartle vacuum'', describing a horizon in thermal equilibrium, of a theory of free Dirac fermions in two spacetime dimensions. In this section, we define our terms and outline this problem in detail.

\subsection{Free Dirac Fields in Curved Spacetime}
We work in a two-dimensional Lorentzian manifold $(\mathcal{M}, g_{\mu \nu}(t, x))$. Every two-dimensional spacetime is conformally flat \cite{CarrollGR}, and thus can be covered by finite-size patches in which the metric $g_{\mu \nu}(t, x)$
differs from the Minkowski metric $\eta_{\mu \nu}\equiv \mathrm{diag}\begin{pmatrix} -1, 1 \end{pmatrix}$ by an overall function $\Omega^2(t, x)$, called the conformal (Weyl, scaling) factor. 
We will always adopt coordinates in which this is manifestly true,
\begin{equation}
\label{eq:ConformalFlat}
g_{\mu \nu}(t, x) = \Omega^2(t,x) \eta_{\mu \nu}.
\end{equation}
The strategy to be outlined will not qualitatively depend upon this choice, though the specific equations we present will. 

The manifold will shelter a ``Dirac'' field of two-component spinors $\psi_A(t, x)$, whose spinor components are indexed with
capital-Latin letters. In order to simplify the notation, we will usually suppress these. When spinor indices are suppressed $(\psi_A(t, x) \to \psi(t, x)$ and $\psi^{\dag A}(t, x) \to \psi^\dag(t, x))$, $\psi(t, x)$ and $\psi^\dag(t, x)$ are to be formally treated, apart from their transformation properties, respectively as two-component column and row vectors.

The spinors form a separate spinor representation of the Lorentz group at each point in $\mathcal{M}$.
In curved spacetime \cite{SupergravityBook, ParkerAndToms}, along with the usual ``flat spacetime'' gamma matrices $\gamma^\mu$ we need ``curved spacetime'' equivalents $\tilde{\gamma}^\mu (t, x)$, respectively defined by their anticommutation relations,
\begin{subequations}
\label{eq:GammaAnticommutation}
\begin{align}
\label{eq:GammaFlatAnti}
\{ \gamma\indices{^{\mu}_{A}^{B}}, \gamma\indices{^{\nu}_{B}^{A}} \} &= 2\eta^{\mu \nu}, \\
\label{eq:GammaCurvedAnti}
\{ \tilde{\gamma}\indices{^{\mu}_{A}^{B}}(t, x), \tilde{\gamma}\indices{^{\nu}_{B}^{A}}(t, x) \} &= 2g^{\mu \nu}(t, x). 
\end{align}
\end{subequations}
When spinor indices are suppressed $(\gamma\indices{^{\mu}_{A}^{B}} \to \gamma{^{\mu}}$,  $\tilde{\gamma}\indices{^{\mu}_{A}^{B}} \to \tilde{\gamma}{^{\mu}})$, $\gamma{^{\mu}}$ and $\tilde{\gamma}{^{\mu}}$ interact with the Dirac spinors as $2\times2$ matrices. In the manifestly conformal chart \eqref{eq:ConformalFlat}, we will choose 
\begin{equation}
\label{eq:GammaDef}
\tilde{\gamma}^\mu(t, x) = \Omega^{-1}(t, x) \gamma^\mu,
\end{equation}
so that \eqref{eq:GammaFlatAnti} imply \eqref{eq:GammaCurvedAnti}.
It is convenient to define the Dirac adjoint,
\begin{equation}
\bar{\psi}(t, x) \equiv i \psi^\dag(t, x) \gamma^0,
\end{equation}
along with the inaccurately-named ``fifth" gamma matrix,
\begin{equation}
\gamma^5 \equiv i \gamma^0 \gamma^1.
\end{equation}

We will distinguish between two derivative operators. The first is the partial derivative, denoted interchangeably by either the usual $\partial_\mu$ or a comma, \begin{equation}
\partial_\mu T(t,x) \equiv T_{,\mu}(t, x).
\end{equation}
The partial derivative of a tensor generically does not transform as a tensor. To remedy this, one introduces the covariant derivative, denoted interchangeably by either $\nabla_\mu$ or a semicolon,
\begin{equation}
\nabla_\mu T(t, x) \equiv T_{;\mu}(t,x).
\end{equation}
The covariant derivative satisfies the axioms of a derivative operator \cite{WaldGR}, notably including linearity and the product rule.
It differs from the partial derivative by the subtraction of ``connection'' terms, such that the full expression transforms covariantly. The specific connection depends on the transformation properties of the operand. Details are available in e.g. \cite{WaldGR, SupergravityBook}, but in the case of the spinor fields we have
\begin{subequations}
\label{eq:SpinDeriv}
\begin{align}
    \bar{\psi}_{;\mu}(t, x) &= \bar{\psi}_{,\mu}(t, x) - \bar{\psi}(t, x) C_\mu(t, x), && C_0(t, x) = -\frac{1}{2} \frac{\Omega_{,1}(t, x)}{\Omega(t, x)} \gamma^0 \gamma^1 \\
\psi_{;\mu}(t, x) &= \psi_{,\mu}(t, x) + C_\mu(t, x) \psi(t, x), && C_1(t, x) = -\frac{1}{2} \frac{\Omega_{,0}(t, x)}{\Omega(t, x)} \gamma^0 \gamma^1,
\end{align}
\end{subequations}
where the expressions in the right column (but not the left) are obviously specific to the manifestly conformal chart \eqref{eq:ConformalFlat}. We can work out $\tilde{\gamma}\indices{^\mu_{;\nu}}(t, x) = 0$ \cite{SupergravityBook}, which in fact holds generally.

We will sometimes use the ``arrow'' notation
\begin{subequations}
\label{eq:ArrowDerivative}
\begin{align}
\bar{\psi}(t,x) \overset{\leftrightarrow}{\nabla}_\mu \psi(t,x) = \bar{\psi}(t,x) \psi_{; \mu}(t,x) - \bar{\psi}_{;\mu}(t,x) \psi(t,x),\\
\bar{\psi}(t,x) \overset{\leftrightarrow}{\partial}_\mu \psi(t,x) = \bar{\psi}(t,x) \psi_{, \mu}(t,x) - \bar{\psi}_{,\mu}(t,x) \psi(t,x).
\end{align}
\end{subequations}
Applying \eqref{eq:SpinDeriv} we see
\begin{equation}
\label{eq:DiracToPartial}
\bar{\psi}(t, x) \tilde{\gamma}^\mu(t, x) \overset{\leftrightarrow}{\nabla}_\mu \psi(t, x) = \frac{1}{\Omega(t, x)}\left(\bar{\psi}(t, x) \gamma^\mu \overset{\leftrightarrow}{\partial}_\mu \psi(t, x)\right),
\end{equation}
which is specific to 2 dimensions. 

Dynamics will be set by the free Dirac action $S$ and its corresponding Lagrangian density $\mathcal{L}(t, x)$,
\begin{align}
\label{eq:DiracActionCurved}
S[\bar{\psi}(t, x), \psi(t, x)] &= -\int dt dx \mathcal{L}(t, x), \\
\mathcal{L}(t, x) &= \sqrt{-g(t,x)}\bar{\psi}(t, x)\left(\frac{1}{2}\tilde{\gamma}^\mu(t,x) \overset{\leftrightarrow}{\nabla}_\mu - m\right)\psi(t, x), \label{eq:LagrangianDensity}
\end{align}
where $g(t, x)$ is the determinant of the spacetime metric. In the manifestly conformal frame \eqref{eq:ConformalFlat} we have
\begin{equation}
\label{eq:MetricDet}
\sqrt{-g(t, x)} = \Omega^2(t, x).
\end{equation}
In flat spacetime, one more commonly uses a Lagrangian density containing $\gamma^\mu \partial_\mu$, as opposed to the  $\frac{1}{2}\gamma^\mu \overset{\leftrightarrow}{\partial}_\mu$ that appears in ours. The two Lagrangian densities differ only by a total derivative, but the present choice \eqref{eq:LagrangianDensity} is more convenient for our purposes due to \eqref{eq:DiracToPartial}, which lets us write
\begin{equation}
    \mathcal{L}(t, x) = \Omega(t, x) \bar{\psi}(t, x)\left(\frac{1}{2}\gamma^\mu \overset{\leftrightarrow}{\partial}_\mu - \Omega(t, x) m\right)\psi(t, x). \label{eq:LagrangianDensityPartial}
\end{equation}

The canonically conjugate momentum to $\psi(t, x)$, $\pi(t, x)$, is defined by 
\begin{subequations}
\label{eq:CanonicalConjugates}
\begin{align}
    \pi(t, x) &\equiv \frac{\partial \mathcal{L}}{\partial\left(\psi_{,0} (t, x)\right)} = \frac{1}{2} \sqrt{-g(t, x)} \bar{\psi}(t, x) \tilde{\gamma}^0(t, x),  \\
    \label{eq:CanonicalPi}
    &= -\frac{i}{2} \Omega(t, x) \psi^\dag(t, x). 
\end{align}
\end{subequations}
Time evolution (Lie transport) of the fields between successive constant-$t$ hypersurfaces of the spacetime is generated by the Hamiltonian,
\begin{subequations}
\label{eq:ContinuumHamiltonian}
\begin{align}
    H_\psi(t, x) &\equiv \int ~dx~ \left(\pi \psi_{,0} + (\pi \psi_{,0})^\dag - \mathcal{L}(t, x)\right),\\
    &= \int ~dx~ \Omega(t, x) \bar{\psi}(t, x) \left(-\frac{1}{2}\gamma^1 \overset{\leftrightarrow}{\partial_1} + \Omega(t, x) m \right) \psi(t, x), \\
    &= \int ~dx~ \Omega(t, x) \psi^\dag(t, x) \left( -\frac{1}{2} \gamma^5 \overset{\leftrightarrow}{\partial_1} + i \Omega(t, x) m \gamma^0 \right) \psi(t, x) 
\end{align}
\end{subequations}
The stress-energy tensor $T_{\mu \nu}(t, x)$ of the Dirac field is \cite{SupergravityBook}
\begin{equation}
\label{eq:SET}
T_{\mu \nu}(t, x) \equiv \frac{-2}{\sqrt{-g(t, x)}} \frac{ \delta S}{\delta g^{\mu \nu}(t, x)} = \frac{1}{2}  \bar{\psi}(t, x)\left(\tilde{\gamma}_{(\mu}(t, x) \overset{\leftrightarrow}{\nabla}_{\nu)}\right) \psi(t, x) .
\end{equation}
A field's stress-energy tensor measures its response to an infinitesimal diffeomorphism, in the sense that 
\begin{equation}
\label{eq:LieGenerator}
Q(t, x) = \int d\Sigma^\mu(t, x) T_{\mu \nu}(t, x)\xi^\nu(t, x),
\end{equation}
where $d \Sigma^\mu(t, x)$ is the integration measure on a hypersurface $\Sigma$, generates Lie transport of the portion of that field from $\Sigma$ along $\xi^\nu(t, x)$. The Hamiltonian \eqref{eq:ContinuumHamiltonian} can also be derived from \eqref{eq:LieGenerator} by fixing $\Sigma$ to be constant-$t$ hypersurfaces with spatial coordinate $x$ and choosing $\xi^\nu(t, x)$ to point along lines of constant $x$ (see Appendix \ref{HamiltonianAppendix}). Unlike \eqref{eq:ContinuumHamiltonian}, however,  \eqref{eq:LieGenerator} could also be used to propagate the fields along a different foliation than the constant-$t$ one, an issue we will not explore further here.

\subsection{The Quantum Field Theory}
\label{QFTSection}
The theory can now be quantized e.g. canonically, by introducing the canonical anticommutation relations \cite{ParkerAndToms}
\begin{subequations}
\label{eq:Canonical}
\begin{align}
\{ \psi_A (t, x), \pi^B(t, x') \} = i \delta(x' - x) \delta\indices{_A^B},\\
\{ \psi_A(t, x), \psi_B(t, x') \} = \{ \pi^A(t, x), \pi^B(t, x') \} = 0,
\end{align}
\end{subequations}
which require $\psi(t, x)$ to be understood as a field of operators. 
Inserting \eqref{eq:CanonicalPi} into \eqref{eq:Canonical}, we find
\begin{subequations}
\label{eq:CanonicalDag}
\begin{align}
\label{eq:CanonicalPsiDag}
\{ \psi_A (t, x), \psi^{\dag B}(t, x') \} &= \frac{\delta(x' - x) \delta\indices{_A^B}}{2\sqrt{\Omega(t, x') \Omega(t, x)}} ,\\
\{ \psi^{\dag A}(t, x), \psi^{\dag B}(t, x') \} &= 0.
\end{align}
\end{subequations}
Note that the Dirac delta in \eqref{eq:CanonicalPsiDag} makes the powers on the $\Omega(t, x)$ and $\Omega(t, x')$ arbitrary, provided they sum to $-1$. We have chosen the present form to ease discretization later.

We take as given a Hilbert space of quantum states $\ket{\Psi(t)}$,
such that the so-called Hadamard function (which is actually a 2$\times$2 matrix containing 4 functions,  see Appendix \ref{SpinorComponents})
\begin{equation}
\label{eq:Hadamard}
G\indices{^{(1)}_{A}^{B'}}(t, x, t', x') \equiv \frac{1}{2} \left\langle \left[\psi_A(t, x), \bar{\psi}^{B'}(t', x') \right] \right\rangle, 
\end{equation}
is finite, provided the primed and unprimed points not be lightlike separated: notably, provided they are not the same point. The primed index in \eqref{eq:Hadamard} transforms as a spinor with respect to Lorentz transformations at $(t', x')$, rather than at $(t, x)$ as does the unprimed index, but these indices will again usually be suppressed, $G\indices{^{(1)}_A^{B'}}(t, x, t', x') \mapsto G^{(1)}(t, x, t', x')$. The superscript ${}^{(1)}$ is a decorator often used in the literature to distinguish $G^{(1)}(t, x, t', x')$ from other two-point functions such as the Feynman propagator (which will not appear here). 

We will most often be interested in the \emph{equal-time} Hadamard function
\begin{equation}
G^{(1)}(t, x, x') \equiv G^{(1)}(t, x, t, x').
\end{equation}

As will be discussed below, the Hadamard function can be used to regulate and renormalize ``quadratic" operators involving quadratic terms in the
fields. Many interesting operators, including the stress-energy tensor \eqref{eq:SET}, fall in this class. However, the stress-energy tensor must be further manicured even after renormalization in order for its expectation value to meet, for example, the divergence-free condition $\langle T_{\mu\nu}(t, x)\rangle^{;\mu}=0$. This poses a significant, though surmountable \cite{MorettiSET, lewis2019hadamard}, complication that will not be addressed here.  In this study we instead use as prototypes the condensate $\langle C_I(t, x) \rangle$, the pseudo-scalar $\langle C_5(t, x) \rangle$, and the current $\langle j_\mu(t, x) \rangle$,
\begin{subequations}
\label{eq:UnsplitOps}
\begin{align}
 C_I(t, x)  &\equiv m \bar{\psi}(t, x) \psi(t, x),
 && 
 \langle C_I(t, x) \rangle = m \langle \bar{\psi}(t, x) \psi(t, x) \rangle,
 \\
  C_5(t, x)  &\equiv \bar{\psi}(t, x) \gamma^5 \psi(t, x),
 && 
 \langle C_5(t, x) \rangle =  \langle \bar{\psi}(t, x) \gamma^5 \psi(t, x) \rangle,
 \\
 j_\mu(t, x) &\equiv \bar{\psi}(t, x) \tilde{\gamma}_\mu(t, x) \psi(t, x),
&& \langle j_\mu(t, x) \rangle = \left\langle \bar{\psi}(t, x) \tilde{\gamma}_\mu(t, x) \psi(t, x) \right\rangle.
\end{align} 
\end{subequations}
Together, they span the four possible quadratic expressions of the form $\psi^{\dag A}\psi_B$ that can be obtained with components of the Dirac spinors (see Appendix \ref{SpinorComponents}).

\subsection{Point-splitting regularization}

\begin{figure}
\centering
\includegraphics[scale=0.3]{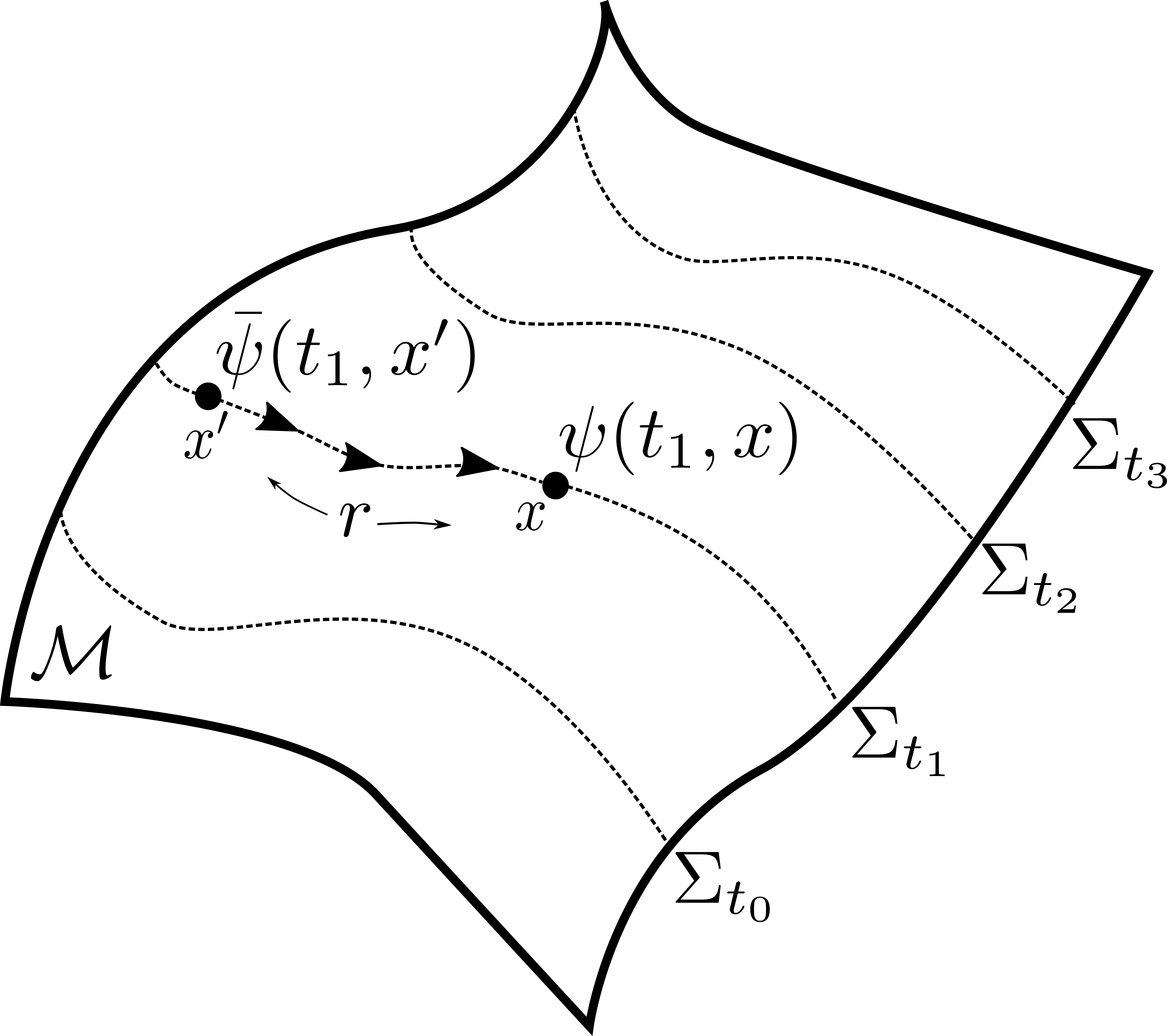}
\caption{Schematic of point-splitting regularization at equal times. The manifold $\mathcal{M}$ is foliated by equal-time surfaces $\Sigma_t$. Instances of quadratic expectation values such as $\langle\bar{\psi}(t, x) \psi(t, x)\rangle$ are replaced with 
coincidence ($r \to 0, \; r \equiv x' - x$) limits of two-point functions such as $\langle\bar{\psi}(t, x') \psi(t, x)\rangle$, expressed in terms of the equal-time Hadamard function $G^{(1)}(t, x, x')$.}\label{fig:PointSplitFig}
\end{figure}

The expectation values on the right column of \eqref{eq:UnsplitOps} are formally infinite
and thus ill-defined. We will replace them with well-defined quantities via so-called ``point-splitting'' \cite{hadamardCauchy, ChristensenReg1978, 1984Ottewill, DEWITT1960, ChristensenVev1976, DecaniniFolacci, AMBRUS2015, lewis2019hadamard}. 
Thus, we regulate our quadratic operators by explicitly treating them as $(t', x') \to (t, x)$ ``coincidence'' limits of traces of the Hadamard function (see Figure \ref{fig:PointSplitFig}). For example, in flat spacetime, we would write
\label{eq:SplitOps}
\begin{eqnarray}
\langle C_I(t, x)  \rangle &\mapsto&  \lim_{(t', x') \to (t, x)} -m \Tr G^{(1)}(t, x, t', x'),\\
\langle C_5(t, x)  \rangle &\mapsto&  \lim_{(t', x') \to (t, x)} -\Tr \gamma^5 G^{(1)}(t, x, t', x'),\\
\langle j_\mu(t, x)  \rangle &\mapsto&  \lim_{(t', x') \to (t, x)} -\Tr \gamma_\mu G^{(1)}(t, x, t', x'),
\end{eqnarray}
where the trace is over the suppressed spinor indices, and the limit is along the unique geodesic connecting $(t', x')$ to $(t, x)$. 

In a curved spacetime, the point-split expressions will
be more complicated, because as e.g. $\bar{\psi}(t', x')$ is parallel-dragged to $(t, x)$, 
it will be deformed and rotated by the spacetime curvature. 
This is expressed symbolically by introducing the ``spin parallel propagator" $\mathcal{J}\indices{_{B'}^A}(t, x, t', x') \mapsto \mathcal{J}(t, x, t', x')$, defined by
\begin{subequations}
\label{eq:Jdef}
\begin{align}
\mathcal{J}_{;\mu}(t, x, t', x') \sigma^{;\mu}(t, x, t', x') &= 0, \\
\lim_{(t', x') \to (t, x)} \mathcal{J}(t, x, t', x') &= \mathbbm{1},
\end{align}
\end{subequations}
where Synge's world function $\sigma(t, x, t', x')$ is one-half the squared geodesic distance from $(t', x')$ to $(t, x)$, while $\mathbbm{1}\indices{_{B'}^A}\mapsto \mathbbm{1}$ is the identity over bispinor indices. As always, the limit is understood to be along the coincident geodesic. These are, in fact, the parallel transport equations,
so that multiplication of $\bar{\psi}(t', x')$ by $\mathcal{J}(t, x, t', x')$ correctly parallel-drags the suppressed spinor indices from $(t', x')$ to $(t, x)$. 

To point split, we now treat the formal expectation values in \eqref{eq:UnsplitOps} as the replacements
\begin{subequations}
\label{eq:CurrentPointSplitReplace}
\begin{align}
\label{eq:CIeqn}
\langle C_I(t, x) \rangle &\mapsto \lim_{(t', x') \to (t, x)} \mathbbm{C}_I(t, x, t', x'), && \mathbbm{C}_I(t, x, t', x') \equiv -m  \Tr \mathcal{J}(t, x, t', x') G^{(1)}(t, x, t', x'),  \\ 
\langle C_5(t, x) \rangle &\mapsto \lim_{(t', x') \to (t, x)} \mathbbm{C}_5(t, x, t', x'), && \mathbbm{C}_5(t, x, t', x') \equiv -\Tr  \mathcal{J}(t, x, t', x') \gamma^5 G^{(1)}(t, x, t', x'),  \\ 
\langle j_\mu(t, x) \rangle &\mapsto \lim_{(t', x') \to (t, x)} \mathbbm{j}_\mu(t, x, t' x'), && \mathbbm{j}_\mu(t, x, t', x') \equiv - \Tr \mathcal{J}(t, x, t', x') \tilde{\gamma}_\mu(t, x) G^{(1)}(t, x, t', x').
\end{align}
\end{subequations}
Notice that, had the spin parallel propagator $\mathcal{J}\indices{_{B'}^{A}}(t, x, t', x')$ not been included in these expressions, the respectively unprimed and primed spinor indices of $G\indices{^{(1)}_{A}^{B'}}(t, x, t', x')$ would transform at different points, and thus could not be traced over to yield a scalar.

To get expressions we can compare to lattice data, we need to specialize each of  \eqref{eq:CurrentPointSplitReplace} to the manifestly conformal metric \eqref{eq:ConformalFlat} with $t = t'$. We will achieve this by expanding $\mathcal{J}(t, x, t, x')$ in the spatial coordinate separation $r \equiv x' - x$. As we will see in \eqref{eq:HadCoord}, the strongest divergences in  \eqref{eq:CurrentPointSplitReplace} are proportional to $1/r$. Thus recovery of the coincidence limit requires this expansion to linear order in $r$. In Appendix \ref{PointSplitAppendix} we find
\begin{equation}
    \label{eq:JExpansion}
    \mathcal{J}(t, x, t, x') = \mathbbm{1}\left(1 - \frac{1}{2}\frac{\Omega_{,0}(t, x)}{\Omega(t, x)} \gamma^0 \gamma^1 r\right) + \cdots.
\end{equation}

Writing $\mathbbm{C}_I(t, x, x') \equiv \mathbbm{C}_I(t, x, t, x')$, $\mathbbm{C}_5(t, x, x') \equiv \mathbbm{C}_5(t, x, t, x')$ and $\mathbbm{j}_\mu(t, x, x') \equiv \mathbbm{j}_\mu(t, x, t, x')$, we thus have
\begin{subequations}
\label{eq:CurrentPointSplitCurved}
\begin{align}
\mathbbm{C}_I(t, x, x') &=  -m\mathrm{Tr} \left( G^{(1)}(t, x, x')\right) - r ~\frac{i m\Omega_{,0}(t,x)}{2\Omega(t,x)} \mathrm{Tr} \left( \gamma^5  G^{(1)}(t, x, x')\right) + \cdots, \\
\mathbbm{C}_5(t, x, x') &=  -\mathrm{Tr} \left(\gamma^5 G^{(1)}(t, x, x') \right)+ r ~\frac{i\Omega_{,0}(t,x)}{2\Omega(t,x)} \mathrm{Tr}\left( G^{(1)}(t, x, x')\right) + \cdots, \\
\mathbbm{j}_0(t, x, x') &= \frac{-1}{\Omega(t,x)} \mathrm{Tr} \left(\gamma_0 G^{(1)}(t, x, x')\right) - r ~\frac{\Omega_{,0}(t,x)}{2\Omega^2(t,x)} \mathrm{Tr} \left(\gamma_1 G^{(1)}(t, x, x')\right) + \cdots, \\
\mathbbm{j}_1(t, x, x') &= \frac{-1}{\Omega(t,x)} \mathrm{Tr} \left(\gamma_1 G^{(1)}(t, x, x')\right) + r ~\frac{\Omega_{,0}(t,x)}{2\Omega^2(t, x)}  \mathrm{Tr} \left(\gamma_0 G^{(1)}(t, x, x')\right) + \cdots .
\end{align}
\end{subequations} 

\subsection{Hadamard renormalization}

The coincidence limits of these expressions \eqref{eq:CurrentPointSplitCurved} are still undefined, because via $G^{(1)}(t, x, x')$ they contain terms that diverge as $x' \to x$. However, given certain smoothness assumptions, bundled together as the requirement that physical quantum states be ``Hadamard", said divergences are completely determined by the local geometry and the mass. In particular, they are independent of the specific choice of ``Hadamard" quantum state. 

The locally determined contributions to \eqref{eq:CurrentPointSplitCurved}, including all the divergences, made by any Hadamard state can then be computed as the first step of a procedure known as Hadamard renormalization. The next step is to subtract the locally-determined terms from each of \eqref{eq:CurrentPointSplitCurved} before the coincidence limits are taken, thus yielding finite answers. 

We label the locally-determined terms with the superscript ``loc'', e.g. $\mathbbm{j}^\mathrm{loc}_\mu (t, x, x')$.
Subtracting each local contribution
from its corresponding bare two-point function within the limit then yields a convergent
remainder - which we label e.g. $\mathbbm{j}^\mathrm{ren}_\mu (t, x, x')$ - with a finite coincidence limit. Next, we identify each of \eqref{eq:UnsplitOps} with the relevant coincidence limit:
\begin{subequations}
\label{eq:QuadraticRenormalized}
\begin{align}
\langle C_I^\mathrm{ren}(t, x) \rangle &\equiv \lim_{x' \to x} \mathbbm{C}^\mathrm{ren}_I(t, x, x') =  \lim_{x' \to x} \left( \mathbbm{C}_I(t, x, x') - \mathbbm{C}^\mathrm{loc}_I(t, x, x') \right), \\
\langle C_5^\mathrm{ren}(t, x) \rangle &\equiv \lim_{x' \to x} \mathbbm{C}^\mathrm{ren}_5(t, x, x') =  \lim_{x' \to x} \left( \mathbbm{C}_5(t, x, x') - \mathbbm{C}^\mathrm{loc}_5(t, x, x') \right), \\
\langle j_\mu^\mathrm{ren}(t, x) \rangle &\equiv \lim_{x' \to x} \mathbbm{j}^\mathrm{ren}_\mu(t, x, x') = \lim_{x' \to x} \left( \mathbbm{j}_\mu(t, x, x') - \mathbbm{j}^\mathrm{loc}_\mu(t, x, x') \right).
\end{align}
\end{subequations}
The locally-determined terms were computed in \cite{lewis2019hadamard}.
They are 
\begin{subequations}
\label{eq:HadCoord}
\begin{align}
\mathbbm{C}^\mathrm{loc}_I(t, x, x') &=  \frac{m^2}{2\pi} \ln \left( \mu \frac{1}{2}\Omega^2(t, x) r^2\right),\\
\mathbbm{C}^\mathrm{loc}_5(t, x, x') &= 0,\\
\mathbbm{j}^\mathrm{loc}_0(t, x, x') &= \frac{\Omega_{,0}(t, x)}{2 \pi \Omega(t, x)}, \\
\mathbbm{j}^\mathrm{loc}_1(t, x, x') &= -\frac{1}{\pi r} + \frac{\Omega_{,1}(t,x)}{2 \pi \Omega(t,x)}, \label{eq:OneCurrent}
\end{align}
\end{subequations}
where $\mu$ is an undetermined dimensionful parameter that must be ``measured''. We will choose
\begin{equation}
\label{eq:mu0}
 \mu = \mu_0 \equiv \frac{1}{2}m^2 e^{2\gamma_E}
\end{equation}
where $\gamma_E = 0.5772\ldots$ is the Euler-Mascheroni constant. This arbitrary choice cancels a constant term that otherwise appears in a short-distance expansion of $\mathbbm{C_I}(t, x, x')$ (see Appendix D), and is thus implicitly made by the standard practice of normal ordering. If a different choice $\mu'$ were made, our results for the condensate would suffer the replacement
\begin{equation}
    \langle C_I^\mathrm{ren}(t, x) \rangle \mapsto \langle C_I^\mathrm{ren}(t, x) \rangle - \frac{m^2}{2\pi}\ln \left( \frac{\mu'}{\mu_0} \frac{1}{2}\Omega^2(t, x)\right).
\end{equation}
Our results are otherwise fixed by the Hadamard renormalization procedure. It is perhaps worth emphasizing that not all of the terms in \eqref{eq:HadCoord} are divergent. Apart from the choice of $\mu$, however, the finite terms are not arbitrary, as e.g. the expected behaviour of the Hawking-Hartle two-point functions would not obtain without them.

\subsection{Hawking-Hartle Vacua}
To specify a calculation we must identify a quantum state.
In Minkowski spacetime, one ``by default'' considers the Minkowski vacuum, selected for example by the demand that $G^{(1)}(t, x , t', x')$ be
Poincar\'{e} invariant. But there is no general such prescription yielding an interesting state in an arbitrary spacetime. States must be identified 
in a manner suggested by the problem at hand. Our target in this study will be the Hawking-Hartle vacuum, which describes a (Killing) horizon in thermal equilibrium with its environment. 

In spacetimes containing such Killing horizons, the Hawking-Hartle vacuum is, in a free theory, the unique state that is both stationary and possessed of a Hadamard function with no divergences in $x$ independently of $x'$. That such a state appears thermal to certain observers is the essential reason for the Unruh effect, the inflationary power spectrum, and black hole radiance, arguably the ``poster children" of quantum field theory in curved spacetime. Given access to a direct simulation of the Hawking-Hartle vacuum, these effects can correspondingly all be studied, perturbed, and prodded in various ways. One could, for example, excite the state, and then study e.g. real time evolution of the von Neumann entropy \cite{InProgress}. 

In the specific spacetimes we study, the Hawking-Hartle vacuum is often referred to by a specific name. In Minkowski spacetime it is the \emph{Minkowski vacuum}. In the prototypical case of Schwarzschild spacetime, it is the \emph{Hawking-Hartle vacuum}, or sometimes the \emph{Kruskal vacuum}. In de Sitter, it is the \emph{Bunch-Davies} or \emph{Euclidean} vacuum. These and similar comparisons are tabulated in Table \ref{tab:TheTable}.

\begin{table}
\centering
\begin{tabular}{|c||c|c||c|c|}
    \hline
     \bf Spacetime & \bf Global Chart & \bf Hawking-Hartle Vacuum  & \bf Static Chart & \bf Static Vacuum   \\
     \hline
     \rm
     Minkowski & Minkowski & Minkowski vacuum & Rindler (Lass) & Rindler vacuum \\
     \hline
     Schwarzschild & Kruskal & Hawking-Hartle/Kruskal vacuum & Schwarzschild  & Boulware vacuum \\
     \hline
     de Sitter & Global & Bunch-Davies/Euclidean vacuum & Static & Static vacuum \\
     \hline
\end{tabular}
\caption{Names commonly used in the literature for the global chart, static chart, Hawking-Hartle vacuum, and static vacuum in the three bifurcate Killing horizon spacetimes we work with.}
\label{tab:TheTable}
\end{table}

\subsubsection{Bifurcate Killing Horizons}
The Hawking-Hartle vacuum is defined in terms of geometric objects called ``bifurcate Killing horizons". Before reviewing the Hawking-Hartle vacuum itself, we will illustrate the concept of a bifurcate Killing horizon, via  concrete examples in Minkowski, two-dimensional Schwarzschild, and de Sitter spacetime. 

In all three spacetimes, we will first identify \emph{global observers} at constant $X$ in the \emph{global coordinates} $(T, X)$. Next we will identify \emph{static observers} at constant $x$ in the \emph{static coordinates} $(t, x)$. We will use these same labels for the global and static coordinates in all three spacetimes. The corresponding Weyl factors \eqref{eq:ConformalFlat} will be written $\Omega_\mathrm{global}^{i}(T, X)$ and $\Omega_\mathrm{static}^{i}(x)$, where the superscript ${}^{i}$ labels the spacetime. That is, we will write $\Omega_\mathrm{global}^{\mathrm{Mink.}}(T,X)$, $\Omega_\mathrm{global}^{\mathrm{Schw.}}(T,X)$, and $\Omega_\mathrm{global}^{\mathrm{dS}}(T,X)$ for global observers and $\Omega_\mathrm{static}^{\mathrm{Mink.}}(x)$, $\Omega_\mathrm{static}^{\mathrm{Schw.}}(x)$, and $\Omega_\mathrm{static}^{\mathrm{dS}}(x)$ for local observers. Notice the static Weyl factors are, by definition, independent of the static time $t$. 

Bifurcate Killing horizon spacetimes are defined by the presence of a lightlike \emph{Killing horizon} beyond which the static timelike coordinate becomes spacelike, so that the static observers are confined to an exterior \emph{static patch}. The Killing horizon is ``bifurcate" because of the reflection-like symmetry relating the top and right patches to the bottom and left ones. They meet at the so-called \emph{bifurcation point}, where a constant $t$ hypersurface in either static wedge intersects the horizon.

The important geometric features are thus the global observers, the static observers, their confinement by the horizon to the static patch, and the bifurcation point. Now, we will highlight these in concrete examples, starting with Minkowski spacetime.

\begin{figure}
\centering
\includegraphics[width=0.5\linewidth]{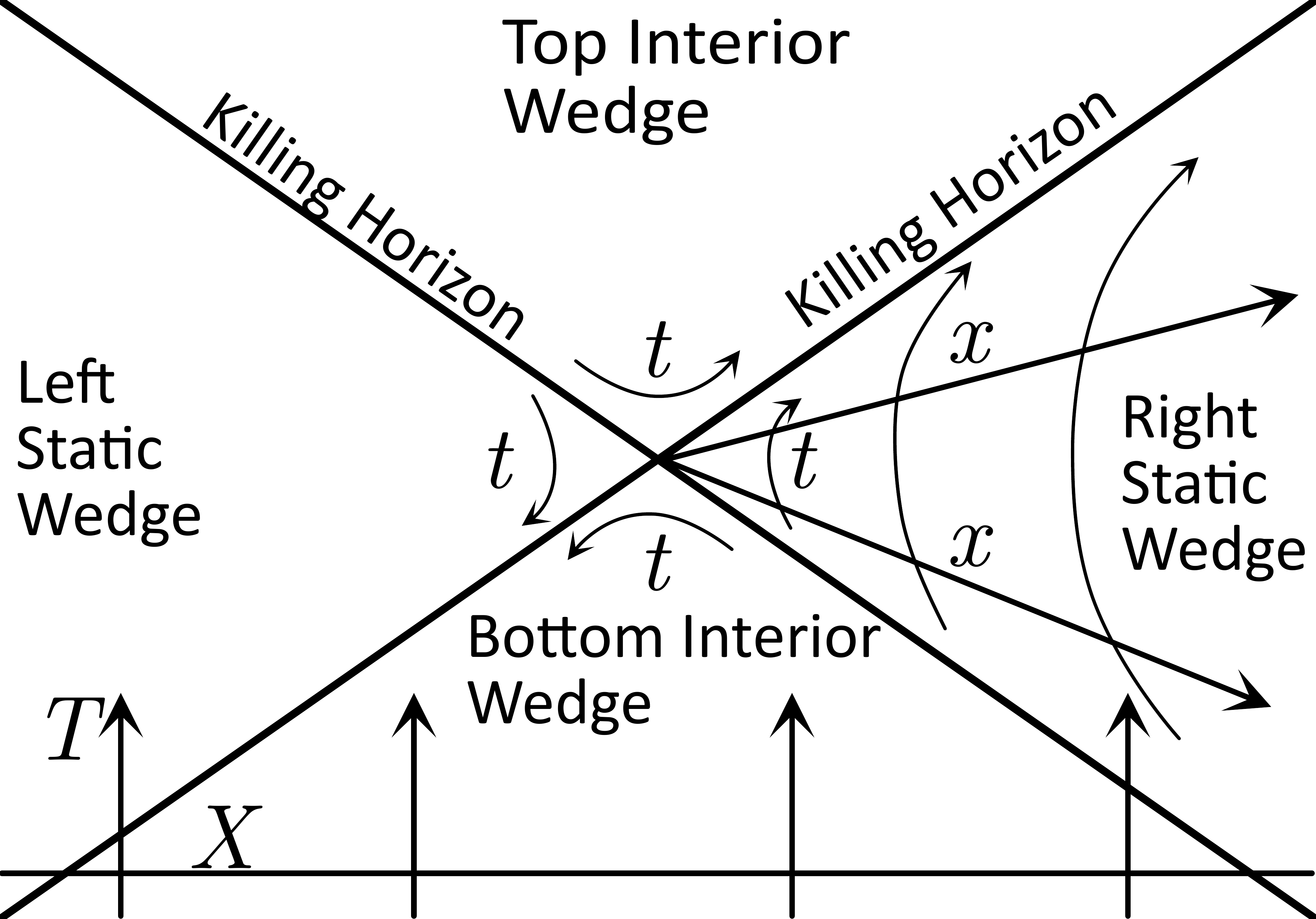}
\caption{Minkowski spacetime, drawn so that inertial (global) coordinates form straight lines. One constant global-space-$X$ surface and a few constant global-times-$T$ observers are depicted as straight horizontal and vertical lines. A particular Killing horizon pair is shown as the diagonal X-shape: it divides Minkowksi spacetime into left and right static wedges, and into top and bottom interior wedges. These horizons are irrelevant to the global observers, but confine the static observers, whose static time $t$ runs hyperbolically in the diagram, to one or the other static wedge. A few static observers are depicted in the right wedge, along with a pair of constant static-space-$x$ hypersurfaces.
}
\label{fig:MinkFig}
\end{figure}

\textit{Minkowski.---} 
The global chart $(T, X)$ on Minkowski spacetime is given by
\begin{equation}
    \label{eq:MinkowskiMetric}
    \Omega_\mathrm{global}^{\mathrm{Mink.}}(T, X) = 1,
\end{equation}
called ``inertial" or ``Minkowski" coordinates.
Figure \ref{fig:MinkFig} depicts Minkowski spacetime aligned with the $T, X$ global coordinates. The Minkowski metric is \emph{static} because it does not depend on $T$. This is a special feature of Minkowski spacetime; i.e. in other examples the global coordinates will not be static. Vectors parallel to $T$ translations form a symmetry of the spacetime, and are thus timelike \emph{Killing vectors}.
Minkowski spacetime has another set of timelike Killing vectors, parallel to boosts. Observers parallel to these are called \emph{Rindler observers}. They are at rest in, for example, the \emph{Lass chart}
\begin{equation}
    \label{eq:LassMetric}
    \Omega^\mathrm{Mink.}_\mathrm{static}(x) = e^{\alpha x},
\end{equation}
which is also static because it does not depend on $t$.

The Lass coordinate $x$ ranges from $-\infty$ to $\infty$, but only covers the right static wedge in the diagram \ref{fig:MinkFig}, where the Killing vector is future-pointing and timelike. In the left static wedge, it is past-pointing and timelike. In the top or bottom wedge, it is spacelike. At the boundary, the Killing horizon, it is lightlike. 

The Killing horizon causally disconnects a particular group of timelike Killing observers from half of Minkowski spacetime. Which group depends on which point of Minkowski spacetime is identified with the origin, and upon the choice of acceleration parameter $\alpha$. This parameter defines the rate at which accelerations diverge as observers closer to the horizons are considered, and can be viewed as characterizing the ``strength" of the horizon. This can be quantified by defining the \emph{surface gravity} $\kappa$,
\begin{equation}
\label{eq:SurfaceGravity}
    k^a k^b_{;a} = \kappa k^b,
\end{equation}
where $k^a$ is the Killing vector, and the expression is to evaluated on the horizon. In Minkowski spacetime we have
\begin{equation}
    \kappa_{\mathrm{Mink.}} = \alpha.
\end{equation}

\textit{Schwarzschild.---} Now consider the Schwarzschild black hole, described by the familiar metric
\begin{equation}
    \label{eq:SchwarzschildSchwarzschild}
    ds^2 = -\left(1-\frac{2M}{\xi}\right)dt^2 + \left(1-\frac{2M}{\mathcal{\xi}}\right)^{-1}d\mathcal{\xi}^2.
\end{equation}
We use the term ``Schwarzschild" black hole to describe the 2D manifold partially covered by \eqref{eq:SchwarzschildSchwarzschild}, not the 4D solution to the Einstein equations. Since this line element is independent of $t$, the constant-$\xi$ ``Schwarzschild" observers are timelike Killing observers. The spatial coordinate $\xi$ ranges from $2M$ (the event and Killing horizon) to $\infty$.
\begin{figure}
\centering
\includegraphics[width=0.5\linewidth]{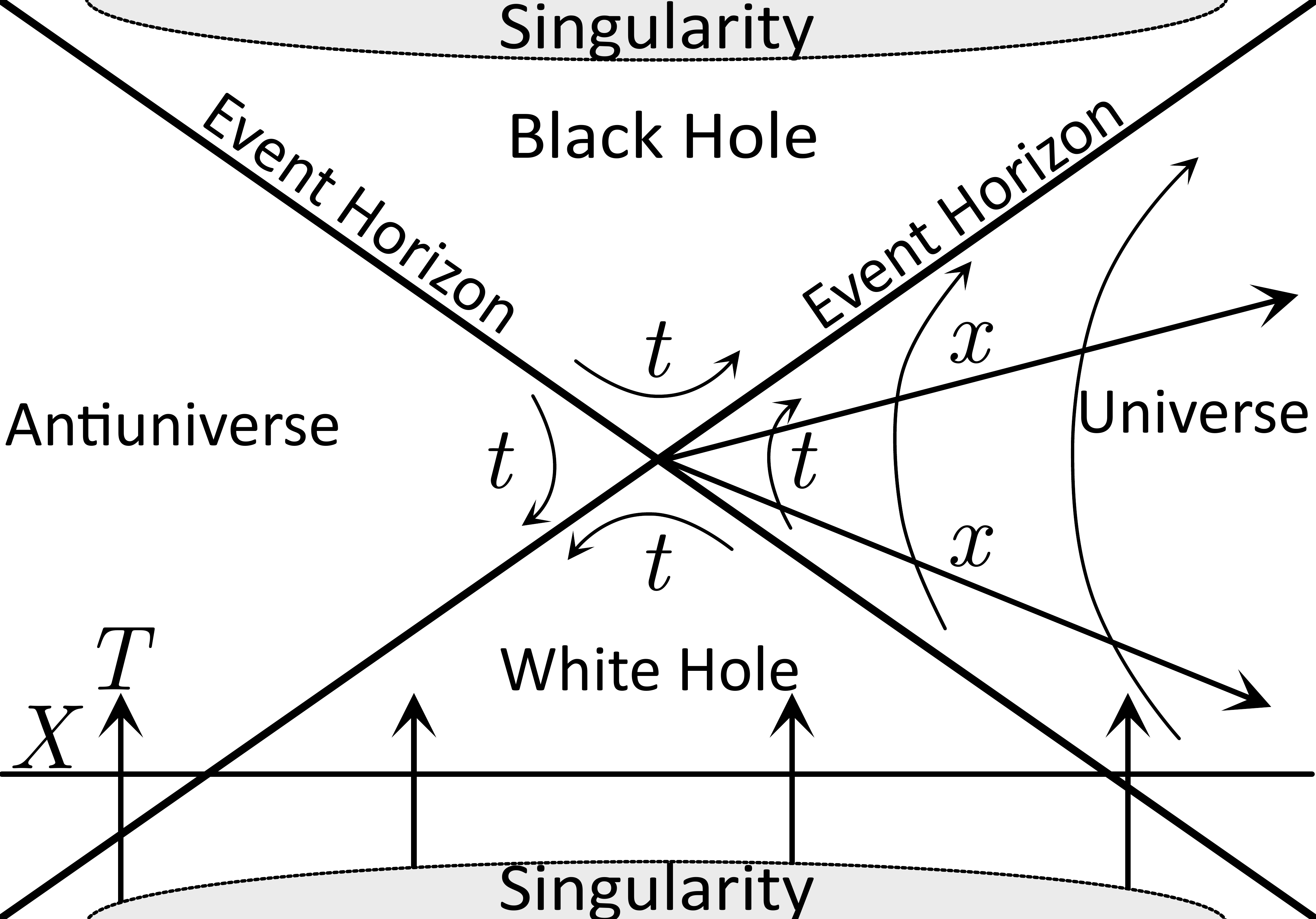}
\caption{Schwarzschild spacetime, drawn so that Kruskal (global) coordinates form straight lines. As in Minkowski spacetime, one constant global-space-$X$ surface and a few constant global-times-$T$ observers are depicted as straight horizontal and vertical lines. In this case the spacetime is divided into four wedges by the event horizon: these wedges are the antiuniverse and universe, analogous to Minkowski spacetime's left and right static wedges, along with the black and white holes, analogous to Minkowski spacetime's top and bottom interior wedges. A few static ``Schwarzschild" observers are depicted in the universe, along with a pair of constant static-space-$x$ hypersurfaces. Unlike Minkowski spacetime, Schwarzschild spacetime exhibits a singularity at $x=0$: curves parameterized by $t$ simply terminate here.
}
\label{fig:SCFig}
\end{figure}
An analogous chart to the Lass metric \eqref{eq:LassMetric} is found by defining the tortoise coordinate $x$
\begin{equation}
    \label{eq:TortoiseDef}
    x \equiv \xi + 2M \ln\left(\frac{\xi - 2M}{2M}\right)
\end{equation}
to find
\begin{equation}
    \label{eq:SchwarzschildTortoise}
    \Omega_\mathrm{static}^\mathrm{Schw.}(x) = \left(1 - \frac{2M}{\xi}\right),
\end{equation} 
which is again static. These coordinates cover the same region as the Schwarzschild coordinates, but have infinite range. The horizon is at $x \to -\infty$.

The region of Schwarzschild spacetime outside the $\xi = 2M$ Killing horizon is thus analogous to the static patch of Minkowski spacetime.
This can be made obvious by adopting the ``Kruskal" chart, which covers all of Schwarzschild spacetime just as the inertial chart covered all of Minkowski. The Kruskal chart is defined by
\begin{subequations}
\label{eq:SchwarzschildKruskal}
\begin{align}
    \label{eq:KruskalDef}
    T^2 - X^2 &= \left(1 - \frac{\xi}{2M}\right) e^{\xi/2M}, \\
    \label{eq:KruskalWeyl}
    \Omega_{\mathrm{global}}^\mathrm{Schw.}(T, X) &= \sqrt{\frac{32 M^3}{\xi} e^{-\xi/2M}}.
\end{align}
\end{subequations}

The const-$\xi$ ``Schwarzschild" observers in the patches are Killing, but they are not geodesic. Their local acceleration diverges towards the horizon at a rate set by the Schwarzschild mass $M$, which is in this way analogous to $\alpha$ in \eqref{eq:LassMetric}. However, while $\alpha$ parameterized a family of charts upon the same manifold, $M$ parameterizes a family of physically inequivalent manifolds. The surface gravity is
\begin{equation}
    \kappa_{\mathrm{Schw.}} = \frac{1}{4M}.
\end{equation}

\begin{figure}
    \centering
    \includegraphics{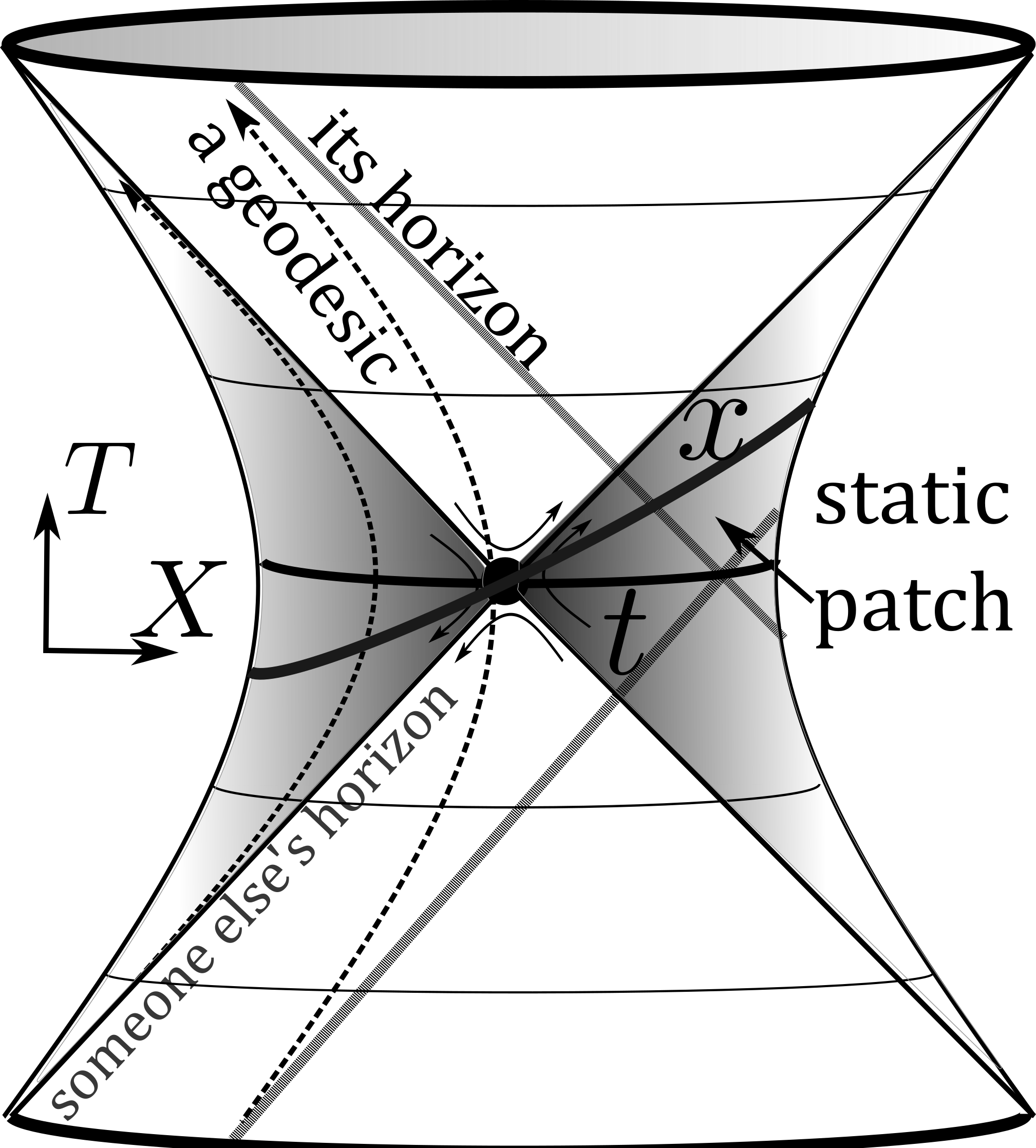}
    \caption{$dS_2$ visualized as an embedding in Minkowski spacetime. The horizontal rings are the constant-global-time $T$ surfaces. Constant global-$X$ geodesic observers, which are the zero-energy geodesics, are constantly boosted with respect to the external flat coordinates, and thus in the embedding diagram follow hyperbolic trajectories like the dotted line. 
    Due to the spacetime expansion, all observers asymptote to such a zero-energy geodesic, which themselves asymptote to lightlike lines, their cosmological horizons. The lightcone of any point on the $T=0$ ring when the expansion reverses forms a bifurcate Killing horizon. The corresponding timelike Killing observers, defining static coordinates $(t, x)$, are not zero energy geodesics, but asymptote to them, as do all worldlines in $dS_2$. Thus, each geodesic's cosmological horizon is also a Killing horizon, confining some set of static observers to some static patch, like the one shaded in the diagram. These static patches also wrap around behind the hyperboloid, and thus form spatially compact diamonds, enclosed on either side.}
    \label{fig:dSdrawing}
\end{figure}

\textit{De Sitter.---} Our final example will be the exponentially expanding de Sitter universe, often labelled $dS_2$ in the two-dimensional case. This spacetime is most straightforwardly discussed in terms of its embedding in higher dimensional flat spacetime, as in Figure \ref{fig:dSdrawing}. The global coordinates $(T, X)$ with line element
\begin{equation}
        \label{eq:dSGlobal}
        ds^2 = -dT^2 + \alpha_{dS}^2 \cosh^2(T/\alpha_{dS}) dX^2
\end{equation}
cover the full hyperboloid, and may be viewed as projections of the higher-dimensional Minkowski coordinates upon it. Constant-$T$ slices are horizontal rings, which contract until $T=0$ and then expand. The ``de Sitter radius" $\alpha_{dS}$ controls the rate of expansion.

Constant-$X$ observers follow timelike geodesics, which appear hyperbolic and thus approach lightlike lines in the embedding diagram, their cosmological horizons. These horizons segment $dS_2$ into causally disconnected patches. In fact, they are Killing horizons, and the geodesic observers are Killing observers, though this is not obvious from \eqref{eq:dSGlobal}. 

We can construct static coordinates bounded by a given Killing horizon as
\begin{equation}
    \label{eq:dSStatic}
    \Omega^\mathrm{dS}_\mathrm{static}(x) = \mathrm{Sech}\frac{x}{\alpha_{dS}}, 
\end{equation}
which cover half the circumference of the central ring. These observers are accelerated relative to the higher-dimensional Minkowski coordinates. The surface gravity is
\begin{equation}
    \kappa_{\mathrm{dS}} = \frac{1}{\alpha_{dS}}.
\end{equation}

\begin{figure}
\centering
\includegraphics[scale=0.8]{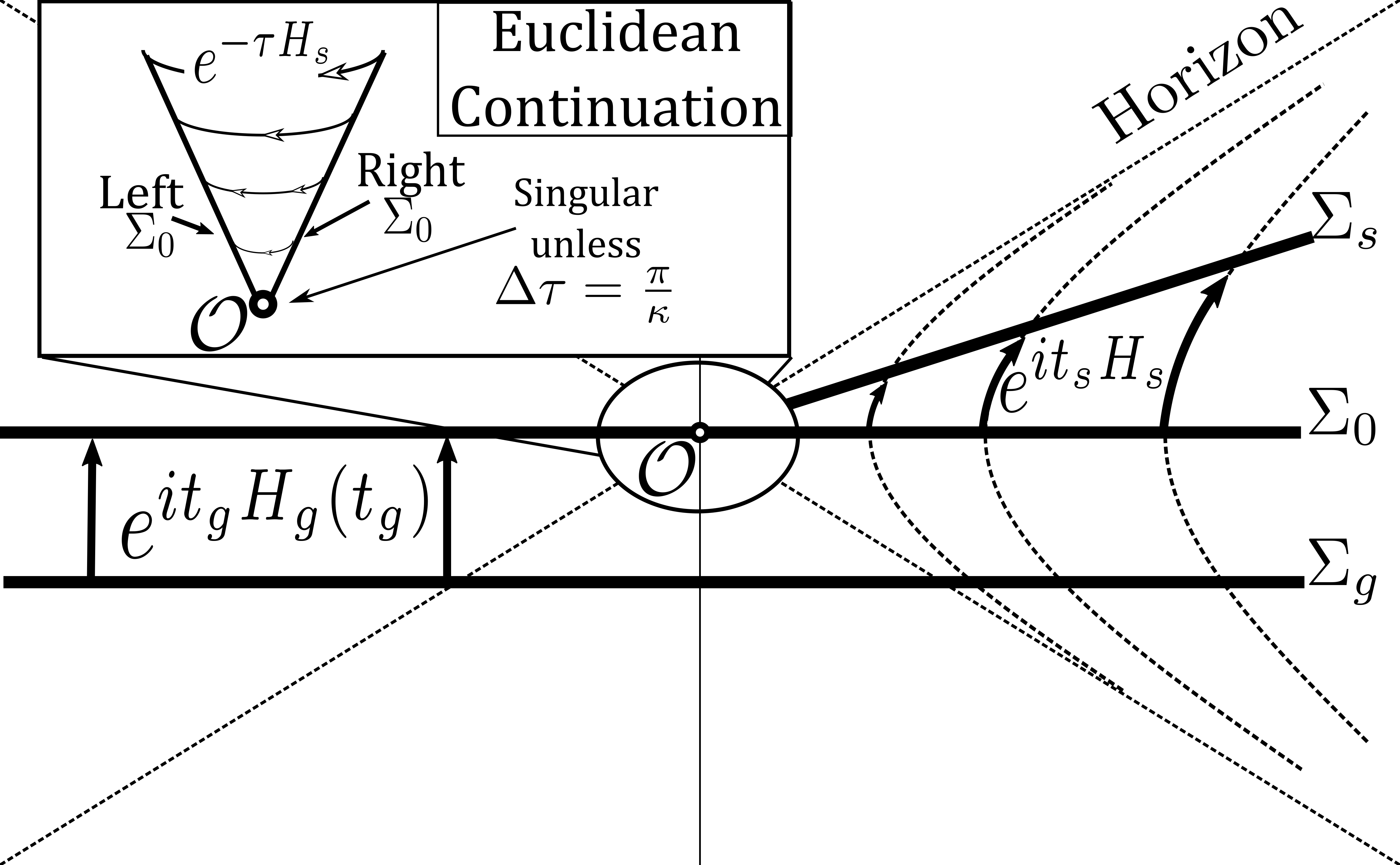}
\caption{Time evolution operators in a bifurcate Killing horizon spacetime. That generated by the global Hamiltonian $H_g(T)$ implements forward
translations in the global time $T$, for example from $\Sigma_g$ to $\Sigma_0$. This translation is always timelike, but it cannot in general be used to define a stationary quantum state because $H_g(T)$ generically depends on the global time. The static Hamiltonian $H_s$ generates forward translations in the static time $t$, for example from the right half of $\Sigma_0$ to $\Sigma_s$. Since $H_s$ is independent of $t$ this translation can be used to define a stationary quantum state, but since the translation is only timelike within the static wedges, that state will generically be singular on the horizon. A nonsingular state that is still stationary with respect to $H_s$ can nevertheless be defined as a Euclidean path integral on the Euclidean manifold found by analytically continuing from e.g. one half of $\Sigma_0$ to the other. This Euclidean manifold is depicted as the half-cone in the inset. The edge of that half-cone overlaps with $\Sigma_0$ in the Lorentzian spacetime, and the path integral is from one edge to the other along the Euclidean time coordinate $\tau$. To fix the singularity at the tip of the cone, $\tau$ must be made periodic with period $\beta = \frac{2\pi}{\kappa}$, resulting in a thermal state of $H_s$ at temperature \protect{\eqref{eq:UnruhTemp}}.}
\label{fig:HHFig2}
\end{figure}

\subsubsection{Hawking-Hartle Vacua}

Now let us try to construct a physically reasonable ``ground" state in a spacetime with a bifurcate Killing horizon. We will make two demands. First, we would like the equal-time Hadamard function to be independent of the static time $t$,  so that we can interpret the state as ``equilibrated". Second, we would like the state not to diverge in any special way at the horizon, since this would be very surprising to the global observers who can cross it. 

These stipulations pick out, uniquely \cite{JacobsonNote} in a free theory, the \emph{Hawking-Hartle vacuum}. To reiterate, the Hawking-Hartle vacuum is the unique state on a bifurcate Killing horizon which is both smooth everywhere and static to the static observers. 

In Figure \ref{fig:HHFig2} we see the time evolution operators generated by two Hamiltonians on the bifurcate Killing horizon spacetime, the static Hamiltonian $H_s$ and the global Hamiltonian $H_g(T)$. These respectively evolve the field forward in the static time $t$ and the global time $T$. The first, $H_s$, evolves the field forward along the worldlines of the static observers - e.g. from the right half of $\Sigma_0$ to $\Sigma_s$ - and is therefore independent of $t$. 

The ``ground state" of $H_s$, called the \emph{static vacuum}, is thus also independent of $t$, our first criterion. However, $H_s$ generates a lightlike rather than timelike translation at the horizon. Because of this, its Hadamard function diverges at the horizon independently of the distance between its arguments. It thus violates our second criterion. 

Though unphysical, the static vacuum comes up often enough to have earned its own special names in each of Minkowski (``Rindler vacuum"), Schwarzschild (``Boulware vacuum") and de Sitter (``static vacuum").

The global Hamiltonian $H_g(T)$ evolves the field forward in the global time $T$, e.g. from $\Sigma_g$ to $\Sigma_0$, which is well behaved at the horizon. The ``ground state" of $H_g(T)$ at a particular time is thus well-behaved there. However, except in Minkowski spacetime, $H_g(T)$ will depend on $T$, and through it, on $t$. This violates our first criterion. Though such a state is potentially physical, it is not ``equilibrated" in any useful sense.

Let us now construct the Hawking-Hartle vacuum, which fulfills both criteria.
Consider a hypersurface such as $\Sigma_0$ in Figure \ref{fig:HHFig2}, formed by the union of constant-time hypersurfaces in the right and left static wedges. A stationary state of $H_s$ prepared everywhere on that hypersurface would be smooth. The difficulties in preparing one stem entirely from the bifurcation point $\mathcal{O}$. 

We will resolve them by analytically continuing past that point. That is, we will make the substitution $t \to \tau = -i t$ in the static chart. Because that chart is by definition time-independent, this only changes the signature of the metric from Lorentzian to Euclidean. For example, the hyperbolic trajectories of the static observers in the Lorentzian manifold become circles in the Euclidean one, as depicted in the inset of Figure \ref{fig:HHFig2}. The ``time" evolution operator $e^{-\tau H_s}$ advances along those circles.

We can follow those circles to move from the part of $\Sigma_0$ in the right static patch to that in the left, without needing to cross the horizon. Of course the Euclidean circles are not physical trajectories, but path integrals along them can be used to define quantum states. These will be stationary with respect to the static Hamiltonian $H_s$, the generator of motion along the circles.

One part of the horizon does lie upon the Euclidean cone: the bifurcation point $\mathcal{O}$. This can be shown to be a literal conical singularity by going to polar coordinates in its vicinity. So long as that singularity persists, the quantum state defined by the path integral on the cone will be singular at the horizon.

Like any conical singularity, however, the singularity can be smoothed by choosing a certain periodicity $\beta_U$ of the relevant angular coordinate, $\tau$,
\begin{equation}
\label{eq:UnruhTemp}
\beta_U = \frac{2\pi}{\kappa},
\end{equation}
the inverse Unruh temperature \cite{PhysRevD.14.870}. States $\rho$ defined as path integrals on closed Euclidean manifolds are thermal with respect to the Hamiltonian generating motion in the direction of the path integral,
\begin{equation}
\rho \equiv \frac{e^{-\beta_U H_s}}{\Tr\left[e^{-\beta_U H_s}\right]}
\end{equation}

Thus the Hawking-Hartle 
vacuum describes the thermal equilibrium state of the horizon. Despite the apparent complexity of the above discussion, preparing it from a Hamiltonian formulation is simple: just find a thermal state of $H_s$ at the relevant Unruh temperature.  

\subsection{Summary}
\label{Procedure}
Everything needed for a well-defined calculation is now at hand. The steps to be followed are:
\begin{enumerate}
\item Choose a spacetime with a bifurcate Killing horizon.
\item Find the Hamiltonian of its static observers by specializing \eqref{eq:ContinuumHamiltonian} to the relevant Weyl factor $\Omega(x)$.
\item Prepare the Hawking-Hartle vacuum as a thermal state of this Hamiltonian at the appropriate Unruh temperature \eqref{eq:UnruhTemp}.
\item Compute the equal-time Hadamard function of that state via \eqref{eq:Hadamard}.
\item Compute renormalized quadratic expectation values via \eqref{eq:QuadraticRenormalized}.
\end{enumerate}
We will next show how to approximately follow this procedure using data obtained by numerically simulating a sequence of lattice theories.

\section{Staggered Fermion Discretization}

One possible strategy to perform numerical simulations is to map the continuum quantum field theory onto a sequence of lattice models. In this section we outline a means to do this, based on a slight adaptation of the well-known ``staggered fermion" discretization \cite{KogutSusskindLattice, Susskind_Lattice_1976, SusskindLattice1976}.

\subsection{Staggered Fermions}
We will numerically simulate the quantum field theory by extrapolating data from a sequence of lattice theories. Thus, consider a one-dimensional lattice whose sites, labelled by $n$, lie at points $x_n = na $ separated by constant lattice spacing $a \equiv x_{n+1} - x_n$. The lattice may be finite or infinite; when finite, it will have an even number of sites $N$. We use a staggered-fermion
discretization \cite{KogutSusskindLattice, Susskind_Lattice_1976, SusskindLattice1976} to relate fermionic operators $\phi_n$ on this lattice to components of the two-component continuum spinors $\psi_A(t_0, x)$ at some reference time $t_0$ 
\begin{subequations}
\label{eq:FermionAnti}
\begin{align}
\{ \phi^\dag_n, \phi_{n'} \} &= \delta_{n, n'}, \\
\{ \phi_n, \phi_{n'} \} &= \{ \phi^\dag_n, \phi^\dag_{n'} \} = 0.
\end{align}
\end{subequations}
As we move from site to site, these operators are identified with alternating components of the continuum spinors,

\begin{subequations}
\label{eq:StaggeredFermions}
\begin{align}
      \psi_0(t_0, x_n) & \mapsto  \frac{1}{\sqrt{2a\Omega(t_0, x_n)}}\phi_n \;\; (n \;\; \mathrm{even}), \\
      \psi_1(t_0, x_n) & \mapsto  \frac{1}{\sqrt{2a \Omega(t_0, x_n)}}\phi_n\;\; (n \;\; \mathrm{odd}).
\end{align}
\end{subequations}
 The normalization $\frac{1}{\sqrt{2a \Omega(t_0, x_n)}}$ recovers the canonical anticommutation relations \eqref{eq:Canonical} and \eqref{eq:CanonicalDag} from the lattice anticommutation relations \eqref{eq:FermionAnti} in the continuum limit $a \to 0$. In addition, it rescales $\phi_n$ to be dimensionless: one of $a$ or $\Omega(t_0, x_n)$ has units of length, depending on whether in $ds^2 = g_{\mu \nu}(t, x) dx^\mu dx^\nu$ the coordinates $dx^\mu$ or the metric components do.

Since this prescription involves individual components of the continuum spinors, we need to commit to a representation of the gamma matrices in order to use it. Our choice will be
\begin{align}
\label{eq:GammaRep}
\gamma^0  = -i \sigma^z, && \gamma^1 = \sigma^y, && \gamma^5 \equiv i \gamma^0 \gamma^1 = -i\sigma^x.
\end{align}
Quadratic operators are transcribed onto the lattice as in \cite{Susskind_Lattice_1976}. Those which couple differing components of $\psi(t, x)$ are mapped to nearest-neighbour couplings such as $\phi^\dag_{n+1} \phi_n$. Those that couple like components are instead mapped to on-site couplings such as $\phi^\dag_n \phi_n$, with oscillating signs introduced while moving from site to site as required by \eqref{eq:GammaRep}.
Noting our slightly different conventions from \cite{Susskind_Lattice_1976}, we have 
\begin{subequations}
\label{eq:StaggeredBilinears}
\begin{align}
    \psi^\dag(t_0, x_n) \gamma^0 \psi(t_0, x_n) &\mapsto \frac{i}{2a \Omega(t, x)} \phi^\dag_n \phi_n  \;\; (n \;\; \mathrm{even}), \\
    \psi^\dag(t_0, x_n) \gamma^0 \psi(t_0, x_n) &\mapsto -\frac{i}{2a \Omega(t, x)} \phi^\dag_n \phi_n  \;\; (n \;\; \mathrm{odd}). 
\end{align}
\end{subequations}

Per \cite{Susskind_Lattice_1976} and our differing normalization of $\phi_n$, the derivative term in the Hamiltonian \eqref{eq:ContinuumHamiltonian} maps to
\begin{equation}
\label{eq:StaggeredDerivative}
   \frac{1}{2} \psi^\dag(t_0, x_n) \overset{\leftrightarrow}{\partial_1} \gamma^5 \psi(t_0, x_n) \mapsto -\frac{i}{2a^2 \Omega(t, x)}\left(\phi^\dag_{n}\phi_{n+1} - \phi^\dag_{n+1} \phi_n \right)
\end{equation}
Comparing \eqref{eq:StaggeredBilinears} and \eqref{eq:StaggeredDerivative} to the continuum Hamiltonian \eqref{eq:ContinuumHamiltonian} yields
\begin{equation}
\label{eq:LatticeHam}
H_\phi(t) = \sum^N_n \left[ \frac{i}{2a}\left(\phi_n^\dag \phi_{n+1} -\phi^\dag_{n+1} \phi_n\right) + (-1)^n \Omega(t, x_n) m \phi^\dag_n \phi_n\right],
\end{equation}
where $dx \mapsto a$ was adopted. 

Thus regulated, such quadratic operators are well-defined on the lattice, but their continuum limits $a \to 0$ will diverge. We extract defined quadratic expectation values by way of the point-split definitions made in Section \ref{QFTSection}. First, we map correlation functions onto the lattice, by first expanding their spinor dependence into components using the representation \eqref{eq:GammaRep}, and then applying \eqref{eq:StaggeredFermions}, as illustrated in Appendix \ref{SpinorComponents}. Starting from
\begin{equation}
\label{eq:TwoPointMap}
\langle \psi^\dag_i(t_0, x') \psi_j(t_0, x) \rangle = \lim_{a \to 0}
    \begin{cases}
        
        \left(2a\sqrt{\Omega(t, x) \Omega(t, x')}\right)^{-1}\left\langle \phi^\dag_n \phi_{n'} \right\rangle,&\; n, i\;\mathrm{same}\; \mathrm{parity}; \; n', j\;\mathrm{same}\; \mathrm{parity} \\
        0 & \mathrm{otherwise}
    \end{cases}
\end{equation}
we make the correspondences
\begin{subequations}
\label{eq:HadamardTrace}
\begin{align}
    \Tr G^{(1)}(t, x, x') &\mapsto \lim_{a \to 0} K_I(t, x, x', a), \nonumber \\
    & K_I(t, x, x', a) \equiv \frac{-(-1)^n}{2a}\left(\frac{\langle\phi^\dag_{n'} \phi_n \rangle}{\sqrt{\Omega(t, x')\Omega(t, x)}} - \frac{\langle \phi^\dag_{n'+1} \phi_{n+1} \rangle}{\sqrt{\Omega(t, x'+a)\Omega(t, x+a)}} \right), \\
    \Tr \gamma^5 G^{(1)}(t, x, x') &\mapsto \lim_{a \to 0} K_5(t, x, x', a) \nonumber\\
    & K_5(t, x, x', a) \equiv i \frac{(-1)^n}{2a} \left(\frac{\langle\phi^\dag_{n'+1} \phi_{n} \rangle}{\sqrt{\Omega(t, x'+a)\Omega(t, x)}} - \frac{\langle \phi^\dag_{n'}\phi_{n+1} \rangle}{\sqrt{\Omega(t, x')\Omega(t, x+a)}}\right),\label{eq:K5eqn}\\
    \Tr \tilde{\gamma}_0(t, x) G^{(1)}(t, x, x') &\mapsto \lim_{a \to 0}K_0(t, x, x', a), \nonumber\\
    & K_0(t, x, x', a) \equiv \frac{i}{2a}  \left(\sqrt{\frac{\Omega(t, x)}{\Omega(t, x')}}     \langle\phi^\dag_{n'} \phi_n \rangle + \sqrt{\frac{\Omega(t, x'+a)}{\Omega(t, x+a)}} \langle \phi^\dag_{n'+1} \phi_{n+1} \rangle \right), \\
    \Tr \tilde{\gamma}_1(t, x) G^{(1)}(t, x, x') &\mapsto \lim_{a \to 0}K_1(t, x, x', a), \nonumber\\
    & K_1(t, x, x', a) \equiv -\frac{i}{2a} \left(\sqrt{\frac{\Omega(t, x)}{\Omega(t, x'+a)}}\langle\phi^\dag_{n'+1} \phi_{n} \rangle + \sqrt{\frac{\Omega(t,x+a)}{\Omega(t, x'+2a)}}\langle \phi^\dag_{n'+2}\phi_{n+1} \rangle \right). \label{eq:K1eqn} 
\end{align}
\end{subequations}

The correlation functions in \eqref{eq:K1eqn} have been chosen so that the total number of lattice units between sites is the same in both terms. This choice is made so that the coincidence ($x' \to x$) limit is correctly renormalized by the locally determined terms \eqref{eq:HadCoord}. It is not necessary to do this in the case of \eqref{eq:K5eqn}, because the coincidence limit is well-defined in that case. Notice also that the coordinate dependencies of the $\Omega(t, x)$ prefactors are located at the same point of the lattice correlator they are proportional to. This is how, for example, the derivative term in \eqref{eq:OneCurrent} gets reproduced on the lattice.

\begin{figure}
\centering
\includegraphics[scale=0.3]{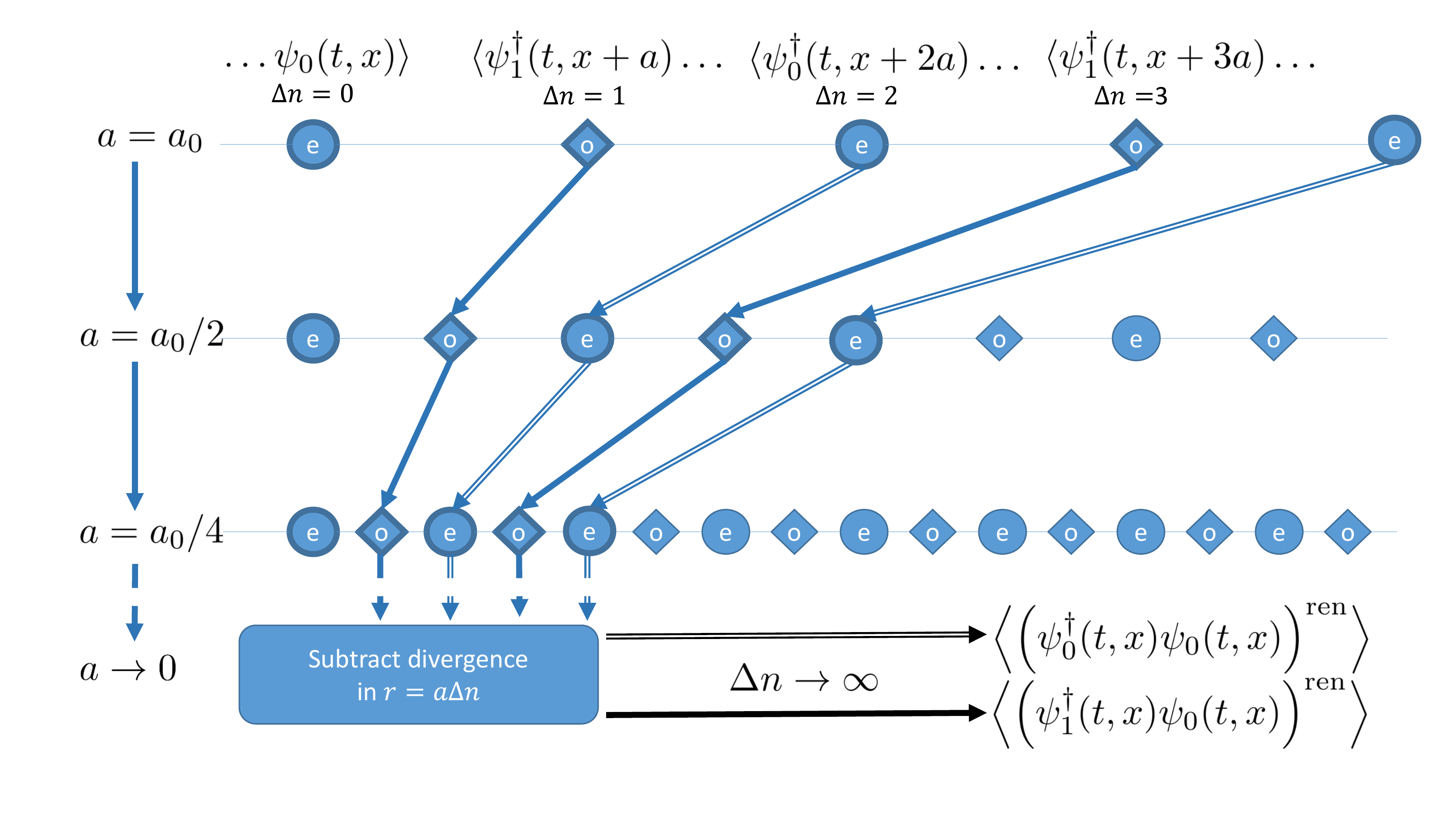}
\caption{Extracting the coincidence limit from lattice correlators. The lattice correlator between sites at $x$ and $x + \Delta n a$, with $\Delta n$ the number of lattice sites separating the two operators and $a$ the lattice spacing, is identified with the continuum correlator $\langle \psi^\dag_i(t, x + \Delta n a) \psi_0(t, x)\rangle$, $i = \Delta n \; \mathrm{mod} \;2$. Holding $\Delta n$ constant and comparing correlators on successively refined lattices, we get functions of $r = \Delta n a$
with the same divergent terms in $r$ as their continuum analogues. Subtracting those divergent terms (in practice by fitting them, up to a constant, to the numerical data) leaves a remainder whose difference from the Hadamard renormalized quadratic expectation value $\langle \psi^\dag_i(t,x) \psi_0(t, x)\rangle^{\mathrm{ren}}$ falls off with $\Delta n$.}
\label{fig:LatticeDrawing}
\end{figure}

As alluded to, the expressions \eqref{eq:HadamardTrace} are finite in the continuum ($a \to 0$) limit, but the continuum quantities they approximate diverge in the subsequent coincidence ($x' \to x$) limit. We can renormalize the latter divergence by subtracting the locally determined terms \eqref{eq:HadCoord} after the continuum limit is taken. In fact, we find empirically that the continuum and coincidence limits can be renormalized simultaneously, by defining
\begin{subequations}
\label{eq:LatticeRenormalized}
\begin{align}
 K_I^\mathrm{ren}(t, x, \Delta n, a) &\equiv K_I(t, x, x+a\Delta n, a) - \frac{m}{2\pi}\ln \frac{1}{2} \mu \Omega^2(x) (a\Delta n)^2, \\
 K_5^\mathrm{ren}(t, x, \Delta n, a) &\equiv K_5(t, x, x+a\Delta n, a),\\
 K_0^\mathrm{ren}(t, x, \Delta n, a) &\equiv K_0(t, x, x+a\Delta n, a) - \frac{\Omega_{,0}(t, x)}{2\pi \Omega(t, x)}, \\
 K_1^\mathrm{ren}(t, x, \Delta n, a) &\equiv K_1(t, x, x+a\Delta n, a) + \frac{1}{\pi (a(\Delta n+1))} - \frac{\Omega_{,1}(t, x)}{2 \pi \Omega(t, x)}.
\end{align}
\end{subequations}
The limit $a \to 0$ with $\Delta n$ held fixed now also brings $x$ and $x'$ together. The remaining terms may depend on $\Delta n$, but do not diverge in the relevant limit $\Delta n \to \infty$. The limiting process is illustrated in Figure \ref{fig:LatticeDrawing}. In practice the $\Delta n$ dependence seems to be quite weak.

We can now make lattice correspondences to the Hadamard renormalized quadratic operators by combining \eqref{eq:LatticeRenormalized} in the $a \to 0$ and $\Delta n \to \infty$ limits with \eqref{eq:QuadraticRenormalized} and \eqref{eq:CurrentPointSplitCurved}, to find 
\begin{subequations}
\label{eq:LatticeQuadraticRenorm}
\begin{align}
    \langle C_I^\mathrm{ren} \rangle (t, x) &= \lim_{\Delta n \to \infty} m\lim_{a \to 0} K_I^\mathrm{ren}(t, x, \Delta n, a), \\
    \langle C_5^\mathrm{ren} \rangle (t, x) &= \lim_{\Delta n \to \infty} \lim_{a \to 0} K_5^\mathrm{ren}(t, x, \Delta n, a),\\
    \langle j_0^\mathrm{ren} \rangle (t, x) &=  \lim_{\Delta n \to \infty} \lim_{a \to 0}\left( K_0^\mathrm{ren}(t, x, \Delta n, a) + \frac{a}{2}(\Delta n+1)  \frac{\Omega_{,0}(t, x)}{\Omega(t,x)} K_1^\mathrm{ren}(t, x, \Delta n, a) \right), \\
    \langle j_1^\mathrm{ren} \rangle (t, x) &=  \lim_{\Delta n \to \infty} \lim_{a \to 0}\left(  K_1^\mathrm{ren}(t, x, \Delta n, a) - \frac{a}{2}(\Delta n+1) \frac{\Omega_{,0}(t, x)}{\Omega(t, x)} K_0^\mathrm{ren}(t, x, \Delta n, a) \right).
\end{align}
\end{subequations}

These expressions are now both well-defined and straightforward to compute numerically. Indeed, we could now make a Jordan-Wigner transformation \cite{JordanWigner} to exactly map the lattice Hamiltonian \eqref{eq:LatticeHam} and lattice Hadamard functions \eqref{eq:HadamardTrace} to expressions in terms of Pauli matrices while retaining their local character. This enables simulations of the ground state of \eqref{eq:LatticeHam} and other related states of interest using matrix product state techniques, as we explore in Ref. \cite{InProgress} in the context of simulating interacting QFTs.

\subsection{The Free Fermion Method}

At present, however, since here we work with a quadratic fermion Hamiltonian $H_\phi(t)$, we can exploit the free fermion formalism to completely and efficiently characterize its (spontaneous) ground state, as well as thermal states and more general Gaussian states, in terms of the equal-time two-point correlation matrix
\begin{equation}
C(t, n, n') = \frac{1}{a} \langle \phi^\dag_n \phi_{n'} \rangle = \frac{1}{a} \Tr (\phi^\dag_n \phi_{n'} \rho(t)),
\end{equation}
where the possible time dependence comes from the Gaussian state $\rho(t)$ under consideration. 

For instance, let us briefly recall how to characterize the instantaneous ground state and thermal states of a more generic quadratic Hamiltonian $H(t) \equiv \sum_{n,k} A_{nn'}(t) \phi_n^{\dagger} \phi_{n'}$  (with $A_{nn'}(t) = A_{n'n}(t)^*$). We proceed by first identifying a unitary transformation $O(t)$ that diagonalizes $A_{nn'}$, that is $O(t)^{\dagger} A(t) O(t) = D(t)$ where $D(t)$ is a diagonal matrix with single-particle energies $\lambda_p(t)$ in its diagonal, that is $D_{pp'}(t) = \delta_{pp'}\lambda_p(t)$. The unitary $O(t)$ in turn defines a unitary $U(t)$ acting on the Hilbert space of the lattice such that it diagonalizes $H(t)$, 
\begin{equation}
U^\dag(t) H(t) U(t) = \sum_p \lambda_p(t) c^\dag_p(t) c_p(t),
\end{equation}
where $c_p(t)$ are a new set of fermionic operators. Then the two point correlator of the Gaussian thermal state with inverse temperature $\beta$,
\begin{equation}
    \rho_\beta(t) \equiv \frac{1}{Z_{\beta}(t)} e^{-\beta H(t)},~~~Z_{\beta}(t) \equiv \Tr e^{-\beta H(t)},
\end{equation}
is given by
\begin{equation}
\left\langle c^\dag_p(t) c_{p'}(t) \right\rangle =\delta_{p, p'} \frac{e^{-\beta \lambda_p(t)}}{1 + e^{-\beta \lambda_p(t)}},
\end{equation}
which implies
\begin{equation}
C(t; n, n') = \frac{1}{a} \langle \phi^\dag_n \phi_{n'} \rangle = \frac{1}{a} \sum_{p,p'} O_{np}(t) \left\langle c^\dag_p(t) c_{p'}(t) \right\rangle O_{p'n'}^{\dagger}(t) =  \frac{1}{a} \sum_p O_{np}(t)O_{n'p}^{*}(t) \frac{e^{-\beta \lambda_p(t)}}{1 + e^{-\beta \lambda_p(t)}}. 
\end{equation}

In our applications below, we will consider the case of a time independent Hamiltonian $H_{\phi}$. Correspondingly, the above expressions simplify to

\begin{eqnarray}
\rho_\beta &\equiv& \frac{1}{Z_{\beta} } e^{-\beta H_{\phi}},~~~Z_{\beta} \equiv \Tr e^{-\beta H_{\phi}},\\
C(n, n') &=& \frac{1}{a} \langle \phi^\dag_n \phi_{n'} \rangle = \frac{1}{a} \sum_p O_{np} O_{n'p}^{*} \frac{e^{-\beta \lambda_p}}{1 + e^{-\beta \lambda_p}}. 
\label{eq:CorrMatUnitary}
\end{eqnarray}

We can now compare \eqref{eq:CorrMatUnitary} with e.g. \eqref{eq:HadamardTrace} to estimate traces of $G^{(1)}(x, x')$. In the simplest case of $\Omega(t, x) = 1$, zero mass $m=0$, zero temperature ($\beta = \infty$) and infinite lattice length $N \to \infty$, we can easily compute the two point lattice correlator exactly to obtain
\begin{equation}
C(n,n') = \frac{i}{2\pi r} \left(-1 + (-1)^{\Delta n}\right),
\end{equation}
where $r \equiv a \Delta n$.
 Comparing with \eqref{eq:HadamardTrace} and then \eqref{eq:CurrentPointSplitCurved}, we see that the Hadamard divergences \eqref{eq:HadCoord} are indeed obtained also on the lattice by sending $a$ to 0 with $\Delta n$ held constant. More generally, we find empirically that this is also true at finite mass $m >0$, finite temperature ($0<\beta<\infty$), and finite lattice length, as well as for space-dependent $\Omega(x)$ (given that the normalizations \eqref{eq:StaggeredFermions} are taken into account).

\section{Applications}

In this section we apply the formalism described above to compute, in the Hawking-Hartle vacua of the two-dimensional Minkowski, Schwarzschild, and de Sitter spacetimes, expectation values of the Hadamard-renormalized quadratic operators defined in \eqref{eq:QuadraticRenormalized}, namely those of the condensate $\langle C_I^\mathrm{ren}(x) \rangle$, the current $\langle j_i^\mathrm{ren}(x)$, and the pseudoscalar $\langle C_5^\mathrm{ren}(x) \rangle$. Note that since all of the metrics we work in will be manifestly static, our notation here differs from that preceding by the suppression of temporal dependencies, e.g. $\langle C_I^\mathrm{ren}(t, x) \rangle \to \langle C_I^\mathrm{ren}(x) \rangle$. 

We will compute the Hawking-Hartle vacuum in four different coordinate charts. First, we will find it in Minkowski spacetime, both as the $\beta \to \infty$ state set by the inertial observers ($\Omega^\mathrm{Mink.}_\mathrm{global} = 1$) and as the $\beta_U = \frac{2\pi}{\alpha}$ state of the Rindler observers, $\Omega^\mathrm{Mink.}_\mathrm{static}(x) = e^{\alpha x}$. We will find equivalent results in both cases despite the quite different lattice Hamiltonians, and in particular will find smooth behaviour when approaching the horizon in the Rindler chart, but only when $\beta = \beta_U$.

Next we will find it in Schwarzschild spacetime, \eqref{eq:SchwarzschildTortoise}. We will find that at the ``Hawking" temperature $\beta_H = 8 \pi M$ results approach those of Minkowski far from the black hole, and yield smooth behaviour approaching it. Finally, we will find the ``Bunch-Davies" vacuum in de Sitter spacetime, $\Omega^\mathrm{dS}_\mathrm{static}(x) = \mathrm{Sech}(x/\alpha_{dS})$, and again will show smooth behaviour approaching the horizon at the Unruh temperature $\beta_U = 2 \pi \alpha_{dS}$.

\subsection{Minkowski: Minkowski Vacuum}
\subsubsection{Inertial Frame}
\begin{figure}
\centering
\includegraphics[width=\textwidth]{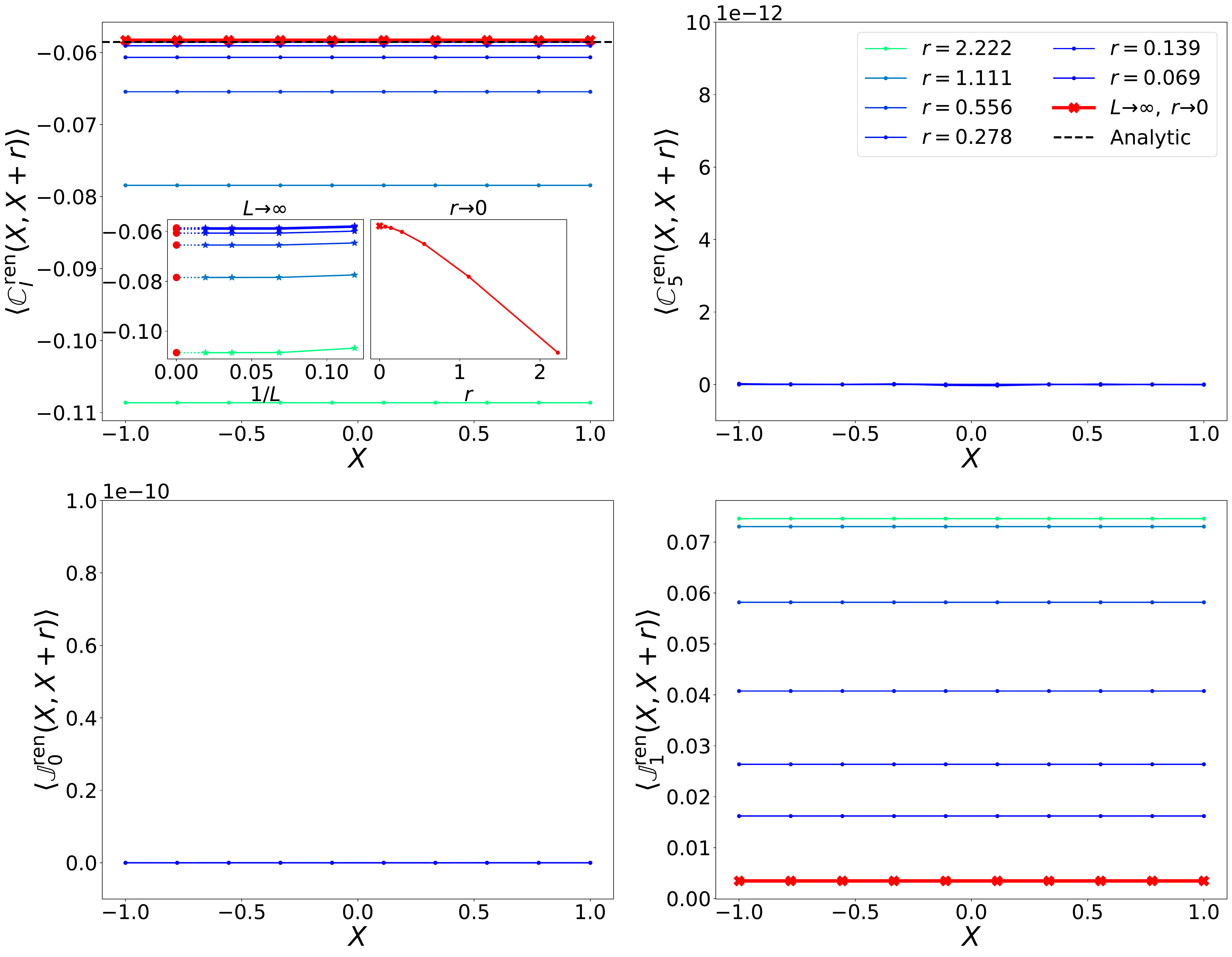}

\caption{Approach to the coincidence limit of the renormalized point-split operators \protect{\eqref{eq:CurrentPointSplitCurved}} in the ground state within the inertial Minkowski chart, computed on a finite and discrete lattice via the discretizations \protect{\eqref{eq:LatticeRenormalized}}, plotted against the inertial coordinate $X$. We use $m=0.6$. Darkening colours indicate simultaneously decreasing point-separation $r$ and lattice spacing $a$. Red lines marked with crosses show the results of the infinite size ($L \to \infty$) and coincidence $r \to 0$ extrapolations, themselves detailed in the insets to the top left and bottom right plot.  $\mathcal{J}_0^\mathrm{ren}(X, X+r)$ (bottom left) and  $\mathcal{C}_5^\mathrm{ren}(X, X+r)$ (top right) vanish essentially to machine precision even at finite separation and lattice size. $\mathcal{J}_1^\mathrm{ren}(X, X+r)$ (bottom right) and $\mathbb{C}_I^\mathrm{ren}(X, X+r)$ (top left) are nonzero on the lattice at finite separation, but vanish in the continuum/coincidence limit.} 
\label{fig:MinkowskiInertialFigs}
\end{figure}
We first discuss the somewhat trivial case of Minkowski spacetime in Minkowski coordinates, $\Omega^\mathrm{Mink.}_\mathrm{global} = 1$. The results in this case (Figure \ref{fig:MinkowskiInertialFigs}) provide a baseline to compare our other results against. Here, we have computed each of \eqref{eq:CurrentPointSplitCurved}, by identifying each with its lattice counterpart in \eqref{eq:LatticeRenormalized}. The superscript ${}^\mathrm{ren}$ indicates that the locally-determined terms \eqref{eq:HadCoord} have been subtracted.

The coloured lines with circular markers show results at finite point-separation, computed on a finite lattice with periodic boundary conditions, after extrapolation to infinite $L$ as detailed in the top inset of the top left panel. We have chosen $m=0.6$ here. The dependence on $X$ is artificial due to the periodic boundary conditions, but we maintain it nevertheless for easier comparison with subsequent results.

Since the locally determined Hadamard terms have been subtracted, the $r \to 0$ extrapolation depicted in the bottom inset of the top left plot in Figure \ref{fig:MinkowskiInertialFigs} yields an estimate of the associated quadratic expectation value in the field theory, $\eqref{eq:QuadraticRenormalized}$. We have not depicted the $\Delta n \to \infty$ limit since it has no visible effect. The results here and throughout are at $\Delta n = 2$.

$\langle C_5 \rangle$ and $\langle J_0 \rangle$ already vanish on the lattice. $\langle J_1 \rangle$, which measures the current density of fermions in the continuum limit, also appears to approach zero, though slowly; the actual extrapolation in this case yields about $10^{-4}$, but can be made smaller by extrapolating from smaller values of $r$. The condensate $\langle C_I \rangle$ vanishes similarly.

\subsubsection{Unruh Effect}
\begin{figure}
\centering
\includegraphics[width=\textwidth]{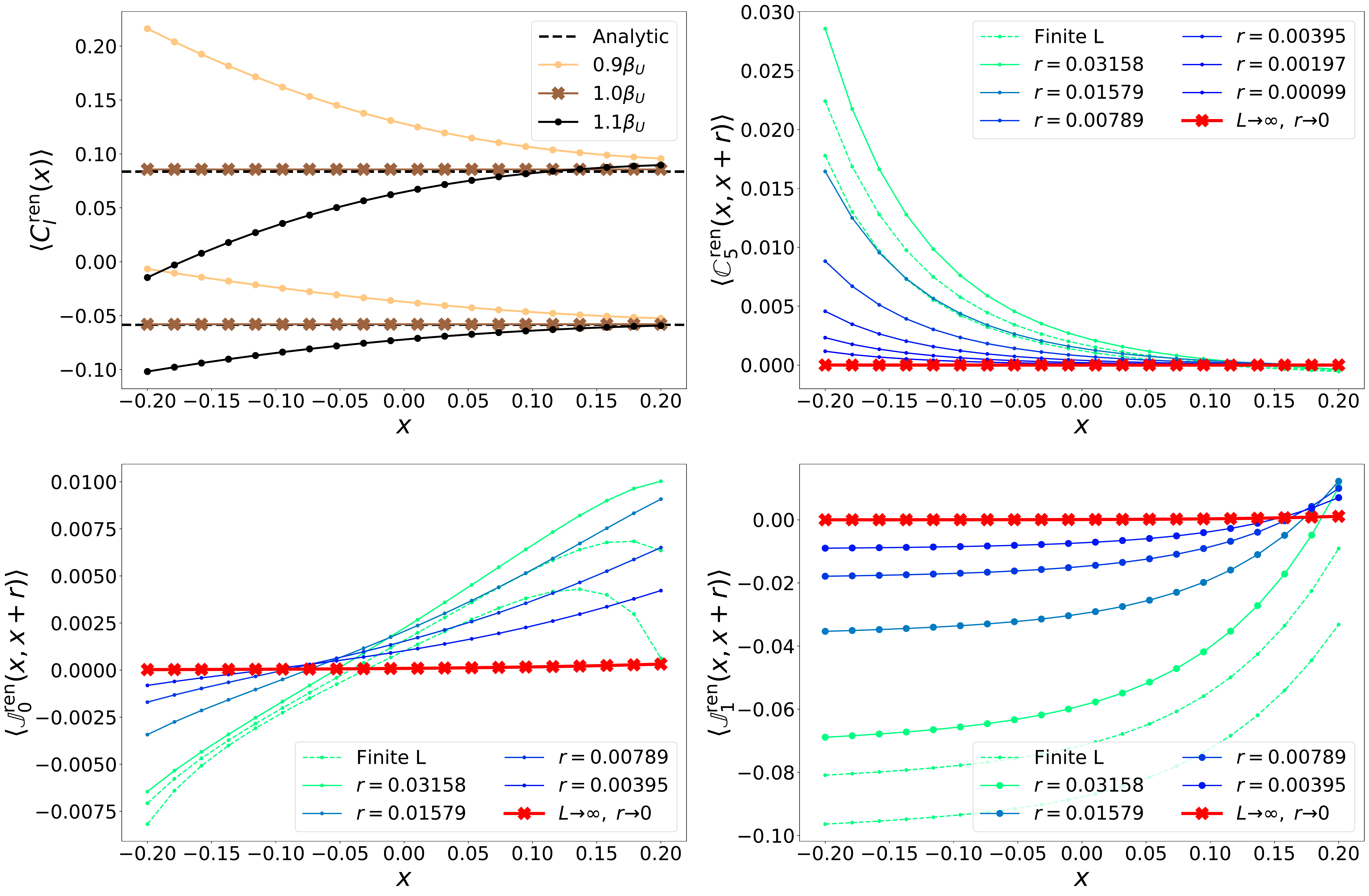}
\caption{Hadamard-renormalized point-split expectation values \protect{\eqref{eq:CurrentPointSplitCurved}} computed, once again in Minkowski spacetime with $m=0.6$, but this time using the Lass chart \protect{\eqref{eq:LassMetric}}.  At $\beta=\beta_U$ this should, in the continuum limit, be the same state as shown in Figure \protect{\eqref{fig:MinkowskiInertialFigs}}; indeed we qualitatively recover those results, in the continuum limit. The top-left plot of $\langle C_I^\mathrm{ren}(x) \rangle$ shows extrapolated results at a few choices of $\beta_U$. The Hawking-Hartle vacuum at $\beta = \beta_U$ is the unique state which does not diverge towards the horizon. It also recovers the inertial values computed in Appendix D, shown as dashed black lines.}
\label{fig:RindlerFigs}
\end{figure}
Thermal states defined by the accelerated Lass chart $\Omega^\mathrm{Mink.}_\mathrm{static}(x) = e^{\alpha x}$ have the same two-point functions as zero-temperature states defined by the inertial Minkowski observers: this is the famous ``Unruh effect". We reproduce this on the lattice by making the same calculations and extrapolations using that metric as we did for the inertial observers. The results are depicted as Figure \ref{fig:RindlerFigs}, where we indeed qualitatively recover the same straight lines as we did in Figure \ref{fig:MinkowskiInertialFigs}. Note that in this case the lattice is not translationally invariant, as reflected in the very different curves away from the continuum limit.

The most interesting example is the condensate $\langle C^\mathrm{ren}_I(x)\rangle$, depicted in the top left plot, since it does not limit to zero. In this case we have plotted results both at the Unruh temperature $\beta_U$ and at temperatures differing from the latter by factors of 0.75 and 1.25, which thus do not yield the Hawking-Hartle vacuum. These, indeed, visibly diverge as the horizon is approached with $x \to -\infty$, while the $\beta_U$ curve remains smooth.

\subsection{Schwarzschild: Hawking-Hartle Vacuum}
\begin{figure}
\centering
\includegraphics[width=\textwidth]{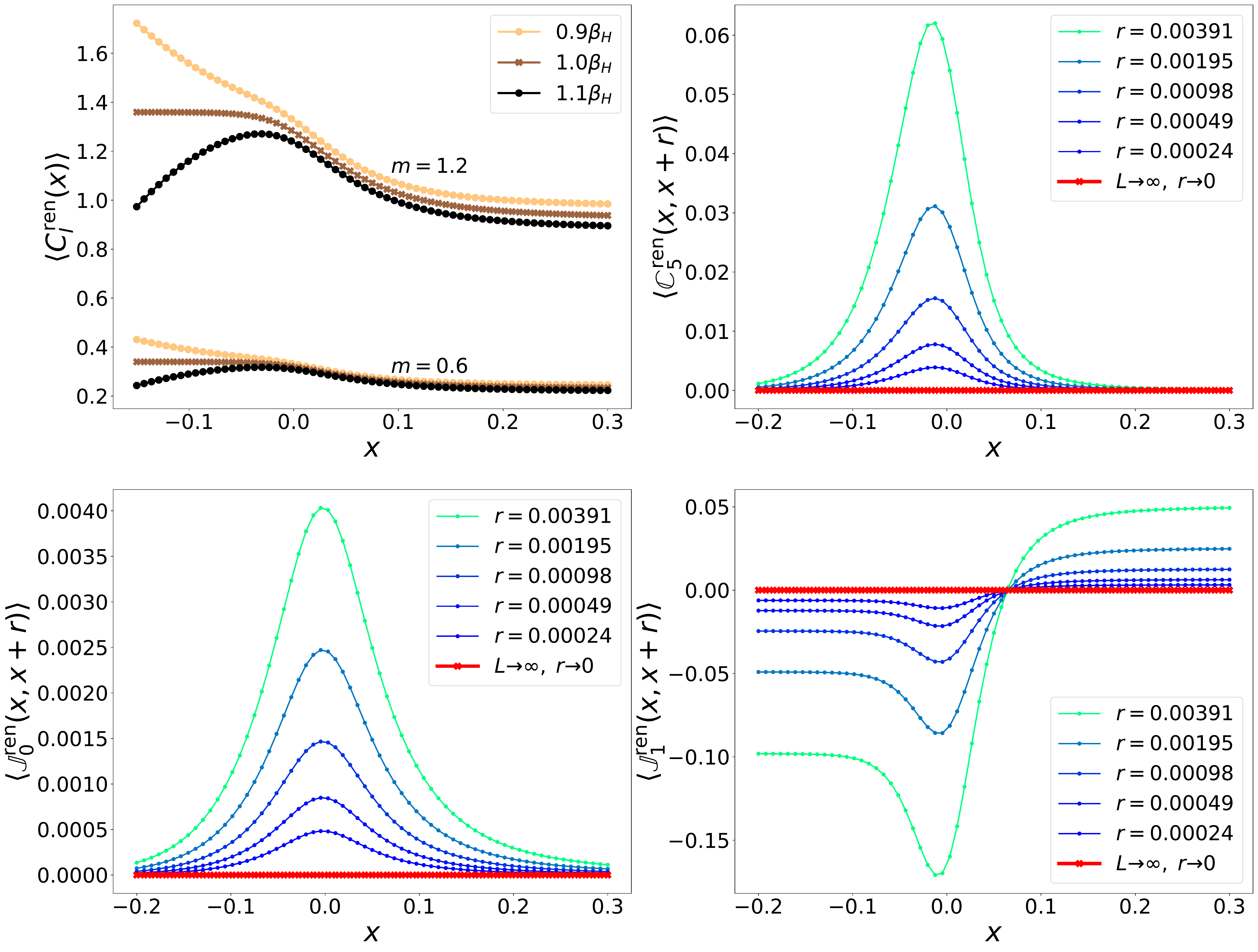}
\caption{ Hadamard-renormalized point-split expectation values \protect{\eqref{eq:CurrentPointSplitCurved}} computed in Schwarzschild spacetime at the Hawking temperaure $\beta_H = 8 \pi M$. All operators except $\langle C_I^\mathrm{ren}\rangle(x)$ fall off to zero in the continuum/coincidence limit. The top-left plot of $\langle C_I^\mathrm{ren}\rangle(x)$ again shows results both at and slightly away from the Hawking increasing temperature $\beta_H$, with brightening colours indicating increasing temperature (and thus decreasing $\beta$). Approaching the black hole, the results away from $\beta_H$ diverge, but those at $\beta_H$ converge to a finite horizon value.}
\label{fig:SchwarzschildFigs}
\end{figure}
Now we repeat the calculation in Schwarzschild spacetime in tortoise coordinates \eqref{eq:SchwarzschildTortoise}. The Hawking-Hartle vacuum is now no longer the Minkowski vacuum, although it approaches the latter far from the black hole $x \to \infty$. Again, all operators but the condensate vanish in the continuum limit, albeit nontrivially (the small bump in $\langle J_0 (x) \rangle$ around $x=0$ shrinks with the smallest value of $r$ used in the extrapolation). Note the full current is conserved, $\langle J_\mu (x) \rangle^{;\mu} = 0$. For a time-independent metric the stipulation that time derivatives ought to vanish as they must in the Hawking-Hartle vacuum, $\langle J_0 (x) \rangle^{,0}$, implies $\langle J_1(x)\rangle = 0$ as well.

The top-left plot again shows the condensate at temperatures at and near the Hawking temperature $\beta_H = 8 \pi M$. Far from the black hole, results approach the inertial Minkowski results at this temperature. Close to it, they approach the Rindler results for a horizon with the same surface gravity. A $x$ around 0, the Hawking-Hartle vacuum smoothly interpolates between these asymptotes.

\subsection{de Sitter: Bunch-Davies Vacuum}
\begin{figure}
\centering
\includegraphics[width=\textwidth]{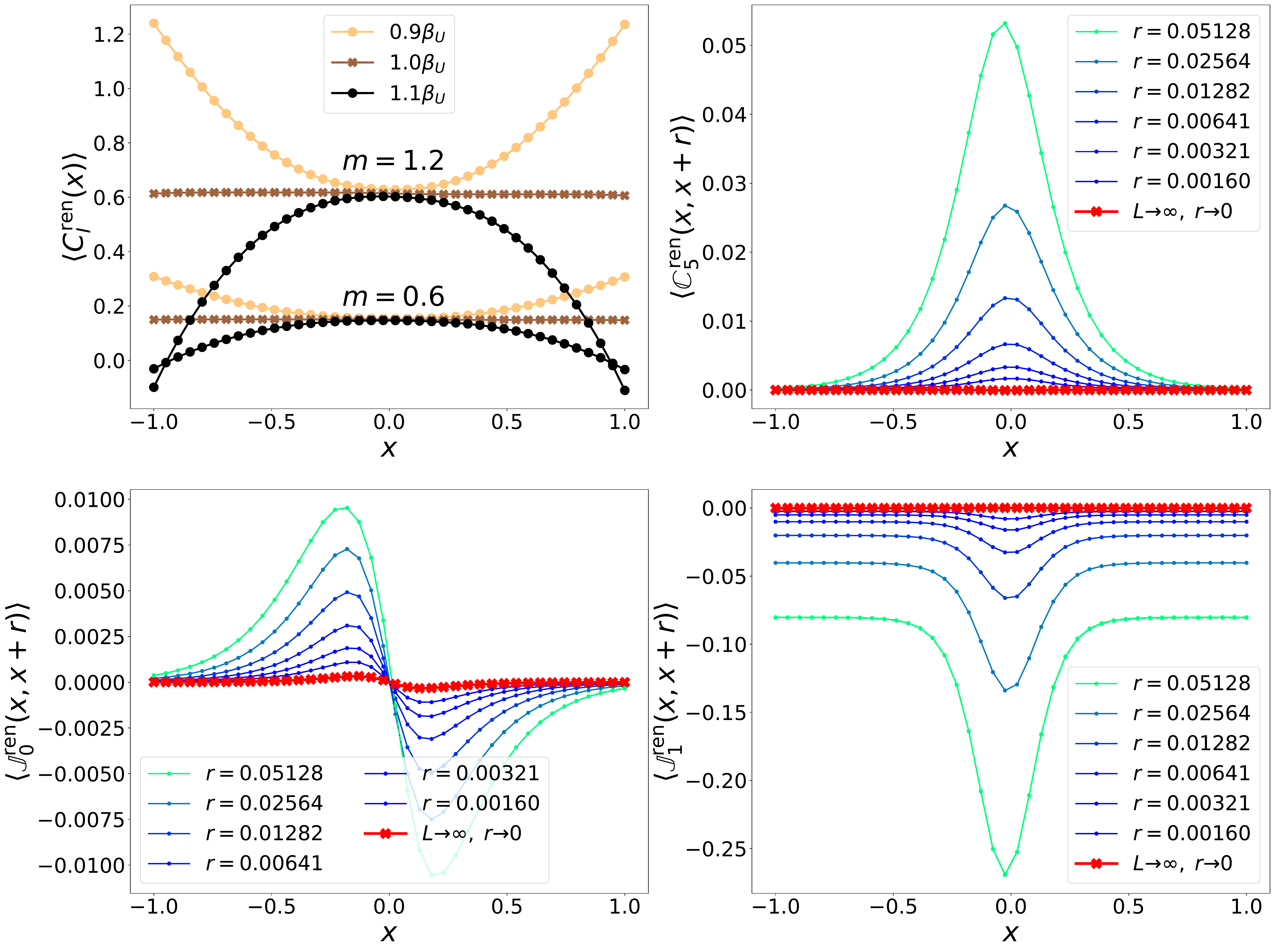}
\caption{Hadamard-renormalized point-split expectation values \protect{\eqref{eq:CurrentPointSplitCurved}} computed in de Sitter spacetime at the Unruh temperaure $\beta_U = 2 \pi \alpha_{dS}$. Here we take $\alpha_{dS} = \frac{1}{2 \pi}$. Once again, all operators except $\langle C_I^\mathrm{ren}\rangle(x)$ fall off to zero, and once again, the top-left plot of the latter shows results both at and slightly away from the Unruh temperature $\beta_U$. Here we used $\delta n = 2$. Notice the slight asymmetry about $x=0$ of the unconverged data, due to $r$ always being added to $x$ (rather than subtracted).}
\label{fig:dSFigs}
\end{figure}
We finally turn to the exponentially expanding de Sitter universe, whose static patch is covered by $\Omega^\mathrm{dS}_\mathrm{static}(x) = \mathrm{Sech}{\frac{x}{\alpha}}$. The Hawking-Hartle vacuum in this case is called the ``Bunch-Davies vacuum". Again, everything except the condensate vanishes in the field theory limit, whereas the condensate smoothly approaches the horizon only at the Unruh temperature.

\section{Conclusion}
Let us review before parting ways. We have illustrated, by simple computations using free Dirac fermions, a method to estimate two-point functions and quadratic expectation values of quantum expectation values upon curved 1+1 dimensional backgrounds. To get finite expectation values we made use of Hadamard renormalization, which perhaps unexpectedly can be applied directly to lattice data, provided the latter be computed holding the lattice length and separation constant. We used these results to demonstrate the Unruh effect, and to compute expectation values in de Sitter spacetime. 

Our eventual goal is to study time-dependent processes for interacting theories in 3+1, which presents three separate avenues for future study. Moving to 3+1 requires us to consider more complicated lattices and to use a different set of Hadamard terms. The numerical study of time-dependent processes is methodologically similar to that of finite temperature states, but doing so in curved spacetime raises the issue of what evolution equation to use. If the Schr\:odinger equation is construed as a projection of the Wheeler-de Witt equation upon a spacetime foliation, it picks up decohering nonlinear terms which are not present if it is instead taken as a first principle. A numerical study would allow the effects of these terms to be analyzed in detail. 

The presence of interactions raises two issues. In this case the free fermion numerical method is no longer available, so we must resort to more complicated, though well-understood, tensor network techniques, which are especially efficient in one spatial dimension. A more serious conceptual difficulty arises when attempting to take the continuum limit. In an interacting theory the appropriate operator weights that terms in the Hamiltonian must incur to yield lines of constant physics are difficult to predict a priori. We mean to study this issue in a future paper.

\section{Acknowledgments}
A. G. M. Lewis is supported by the Tensor Network Initiative at Perimeter Institute. 
Research at Perimeter Institute is supported by the Government of Canada through the Department of Innovation, Science and Economic Development Canada and by the Province of Ontario through the Ministry of Research, Innovation and Science.
G. Vidal is a CIFAR fellow in the Quantum Information Science Program. 
X is formerly known as Google[x] and is part of the Alphabet family of companies, which includes Google, Verily, Waymo, and others (www.x.company). 

\appendix
\section{Coordinate expansion of spin parallel propagator}
\label{PointSplitAppendix}
In this Appendix we find \eqref{eq:JExpansion} by expanding the spin parallel propagator $\mathcal{J}(t, x, t', x')$ defined by \eqref{eq:Jdef} to linear order in $r \equiv x' - x$ with $t' = t$ in the manifestly 
conformal metric \eqref{eq:ConformalFlat}. Writing the Ansatz
\begin{equation}
\label{eq:JAnsatz}
\mathcal{J}(t, x, t, x') = \mathbbm{1} + j(t, x) r + \ldots,
\end{equation}
this task amounts to finding $j(t, x)$. The partial
derivatives of \eqref{eq:JAnsatz} in a coordinate basis are
\begin{subequations}
\label{eq:JPD}
\begin{align}
    \mathcal{J}_{,0}(t, x, t, x') &= j_{,0}(t, x) r, \\
    \mathcal{J}_{,1}(t, x, t, x') &= j_{,1}(t, x) r - j(t, x).
\end{align}
\end{subequations}
In this Appendix we will adopt the standard practice of denoting coincidence limits with square brackets, e.g.
\begin{equation}
\left[ \mathcal{J}(t, x, t', x') \right] = \lim_{(t', x')\to(t, x)} \mathcal{J}(t, x, t', x'),
\end{equation}
where as always the limits are understood to be along the coincident geodesic. Applied to \eqref{eq:JPD} we find
\begin{equation}
    \label{eq:JPDcoinc}
    \left[ \mathcal{J}_{,\mu}(t, x, t, x') \right] = -\delta_{\mu, 1} j(t, x).
\end{equation}
On the other hand, using the spin covariant derivatives listed in \eqref{eq:SpinDeriv}, we find
\begin{equation}
\label{eq:ParallelPropDeriv}
\left[ \mathcal{J}_{;1}(t, x, t', x') \right] = \left[\mathcal{J}_{,1}(t, x, t', x') \right] - \frac{1}{2} \frac{\Omega_{,0}(t, x)}{\Omega(t, x)} \gamma^0 \gamma^1.
\end{equation}
Taking the covariant derivative of the definition \eqref{eq:Jdef}, we next find
\begin{equation}
    \label{eq:JdefDeriv}
    \mathcal{J}_{;\mu\nu}(t, x, t', x')\sigma^{;\mu}(t, x, t', x') + \mathcal{J}_{;\mu}(t, x, t', x') \sigma\indices{^{;\mu}_{\nu}}(t, x, t', x') = 0
\end{equation}
Coincidence limits of $\sigma(t, x, t', x')$ and its derivatives are documented in many sources, e.g. \cite{DEWITT1960, ChristensenVev1976, Poisson2011}. The ones we need are
\begin{subequations}
\label{eq:SigmaCoinc}
\begin{align}
    \left[\sigma^{;\mu}(t, x, t', x')\right] &= 0, \\
    \left[\sigma\indices{^{;\mu}_{\nu}}(t, x, t', x')\right] &= g\indices{^{;\mu}_{\nu}}(t, x).
\end{align}
\end{subequations}
In light of \eqref{eq:SigmaCoinc}, taking the coincidence limit of both sides of \eqref{eq:JdefDeriv} yields 
\begin{equation}
    \label{eq:JderivCoinc}
    \left[ \mathcal{J}_{;\nu}(t, x, t', x') \right] = 0.
\end{equation}
Combining \eqref{eq:JderivCoinc}, \eqref{eq:JPDcoinc}, and \eqref{eq:JdefDeriv} we find
\begin{equation}
    j(t, x) = -\frac{1}{2} \frac{\Omega_{,0}(t, x)}{\Omega(t, x)} \gamma^0 \gamma ^1 \mathbbm{1}
\end{equation}
which, inserted back into \eqref{eq:JAnsatz}, yields the desired result \eqref{eq:JExpansion}.

\section{ADM derivation of Hamiltonian}
\label{HamiltonianAppendix}
In this Appendix we compute \eqref{eq:ContinuumHamiltonian} from \eqref{eq:LieGenerator} and \eqref{eq:SET}. We thus consider a foliation of spacetime by hypersurfaces $\Sigma_t$, upon each of which the spacetime coordinate $t$ is held constant. The hypersurfaces will be charted by the same spatial coordinate $x$ as the full spacetime, see e.g. \cite{ADMpaper, AlcubierreNR, baumgarte2010numerical}. The integration measure on each hypersurface is
\begin{equation}
    d \Sigma^\mu = -dx \sqrt{h(t, x)} n^\mu(t, x),
\end{equation}
where $\nu^\mu$ is the unit ($\nu^\mu \nu_\mu = -1 $) normal vector to the foliation, and $h(t, x)$ is the determinant of the intrinsic metric $h_{\mu \nu}(t, x)$ to the hypersurfaces,
\begin{equation}
h_{\mu \nu}(t, x) \equiv g_{\mu \nu}(t, x) + n_\mu(t, x) n_\nu(t, x).
\end{equation}
We wish to generate Lie transport of the Dirac field in the direction of constant $x$, and thus choose $\xi^\mu(t, x)$ to be the constant coordinate vector, which points along lines of constant $x$ and is normalized by
\begin{equation}
\xi^\mu(t, x) t_{;\mu} = 1.
\end{equation}

Now define the lapse function $\alpha(t, x)$ and shift vector $\beta^\mu(t, x)$ by
\begin{equation}
\xi^\mu(t, x) = \alpha(t, x) n^\mu(t, x) + \beta^\mu (t, x).
\end{equation}
We have
\begin{align}
\label{eq:NormalADM}
n^\mu = (1/\alpha, -\beta^i / \alpha), && n_\mu = (-\alpha, 0),
\end{align}
\begin{equation}
g_{\mu \nu}(t, x) = \begin{pmatrix}-\alpha^2(t, x) + \beta_k(t, x)  \beta^k(t, x) && \beta_i(t, x) \\ \beta_j(t, x) && h_{ij}(t, x) \end{pmatrix}
\end{equation}
where indices $i, j, \ldots$ run over space only. In the manifestly conformal chart \eqref{eq:ConformalFlat} we can read off
\begin{align}
\beta^\mu(t, x) &= 0, && n^\mu(t, x) = (\Omega^{-1}(t, x),\; 0), \\
\alpha(t, x) &= \Omega(t, x), && \xi^\mu(t, x) = (1,\; 0).
\end{align}
Noting
\begin{equation}
\sqrt{h(t, x)} = \Omega(t, x)
\end{equation}
we find from \eqref{eq:LieGenerator}
\begin{equation}
\label{eq:HConformal}
H(t) = -\int dx T_{00}(t, x)
\end{equation}
and from \eqref{eq:SET}
\begin{equation}
\label{eq:T00}
T_{00}(t, x) = \frac{1}{2}\bar{\psi}(t, x) \tilde{\gamma}_0 \overset{\leftrightarrow}{\nabla}_0\psi(t, x).
\end{equation}
In a sense this completes the derivation, but the time derivatives are inconvenient for us. To eliminate them, we need to use the equations of motion (Dirac equations) for $\psi(t, x)$ and $\bar{\psi}(t, x)$, which themselves are found by varying the action \eqref{eq:DiracActionCurved} with respect to the fields. In the presently most convenient form, they are
\begin{subequations}
    \label{eq:DiracTime}
    \begin{align}
        \psi_{;0}(t, x) &= \gamma^0 \gamma^1 \psi_{;1}(t, x) - m \tilde{\gamma}^0(t, x) \psi(t, x) \\
        \bar{\psi}_{;0}(t, x) &= -\bar{\psi}_{;1}(t, x) \gamma^0 \gamma^1 + m \bar{\psi}(t, x) \tilde{\gamma}_0(t, x)
    \end{align}
\end{subequations}
Applying \eqref{eq:DiracTime} to \eqref{eq:T00} yields
\begin{equation}
    \label{eq:T00middle}
    T_{00} = \Omega(t, x) \bar{\psi}(t, x) \left( \frac{1}{2}\gamma^1 \overset{\leftrightarrow}{\nabla}_1 - \Omega(t, x) m \right) \psi(t, x),
\end{equation}
which, in light of \eqref{eq:HConformal}, indeed recovers \eqref{eq:ContinuumHamiltonian}.

\section{Spinor Components}
\label{SpinorComponents}

For the sake of simplicity, in the main text we have mostly suppressed the spinor indices to lighten an otherwise already heavy notation. In this Appendix we will rewrite a few key expressions explicitly in terms of spinor components, given the important role the two spinor components play in the staggered lattice discretization. Recall that we use the gamma matrix representation $\gamma^{0}=-i\sigma^z = -\gamma_0$, $\gamma^1=\sigma^y = \gamma_1$ and $\gamma^5 \equiv i \gamma^0 \gamma^1 = -i\sigma^x$.

In two spacetime dimensions the Dirac field $\psi(t,x)$ is a two-component column spinor
\begin{equation}
\psi(t,x) = \left( \begin{array}{c} \psi_0(t,x) \\ \psi_1(t,x) \end{array} \right)
\end{equation}
whereas its adjoint conjugate $\psi^{\dagger}(t,x)$ and Dirac conjugate $\bar{\psi}(t,x)$ correspond to the two-component row spinors
\begin{equation}
\psi^{\dagger}(t,x) = \left(\begin{array}{cc} \psi^\dagger_0(t,x) & \psi^{\dagger}_1(t,x) \end{array} \right),~~~~\bar{\psi}(t,x) \equiv i \psi^{\dagger}(t,x)\gamma^0 = \left(\begin{array}{cc} \psi^\dagger_0(t,x) & -\psi^{\dagger}_1(t,x) \end{array} \right).
\end{equation}
Then the Hadamard function $G^{(1)}(t, x, t', x')$ is a $2\times 2$ matrix given by
\begin{eqnarray}
\label{eq:Hadamard2}
G^{(1)}(t, x, t', x') &\equiv& \frac{1}{2} \left\langle \left[ \left( \begin{array}{c} \psi_0(t,x) \\ \psi_1(t,x) \end{array} \right), \left(\begin{array}{cc} \psi^\dagger_0(t',x') & -\psi^{\dagger}_1(t',x') \end{array} \right) \right] \right\rangle\\
&=&\frac{1}{2} \left\langle  \left( \begin{array}{cc} 
\left[ \psi_0(t,x), \psi^\dagger_0(t',x')\right] & \left[ \psi_0(t,x), -\psi^{\dagger}_1(t',x')\right] \\ 
 & \\
\left[\psi_1(t,x), \psi^\dagger_0(t',x')\right] & \left[\psi_1(t,x), -\psi^{\dagger}_1(t',x')\right]
\end{array} \right)
\right\rangle.
\end{eqnarray}
It is useful to compute the following traces involving $G^{(1)}(t, x, t', x')$ and $\gamma^5=-i\sigma^x$, $\gamma_0 = i\sigma^z$ and $\gamma_1 = \sigma^y$,
\begin{subequations}
\label{eq:HadamardTrace2}
\begin{align}
    \Tr \left( G^{(1)}(t, x, t', x')\right) &= \frac{-1}{2}\left(  \langle[\psi^\dag_0(t, x'), \psi_0(t, x)]\rangle - \langle [\psi^\dag_1(t, x'), \psi_1(t, x)]\rangle \right), \\
    \Tr \left( \gamma^5 G^{(1)}(t, x, t', x') \right)&= \frac{i}{2} \left( \langle[\psi^\dag_0(t', x'), \psi_1(t, x)]\rangle - \langle [\psi^\dag_1(t', x'), \psi_0(t, x)]\rangle \right),\\
    \Tr \left(\gamma_0 G^{(1)}(t, x, t', x')\right) &= \frac{-i}{2} \left( \langle[\psi^\dag_0(t', x'), \psi_0(t, x)]\rangle + \langle[\psi^\dag_1(t', x'), \psi_1(t, x)] \rangle\right), \\
    \Tr \left(\gamma_1 G^{(1)}(t, x, t', x') \right)&= \frac{i}{2} \left( \langle[\psi^\dag_0(t', x'), \psi_1(t, x)]\rangle + \langle[\psi^\dag_1(t', x'), \psi_0(t, x)]\rangle \right).
\end{align}
\end{subequations}
In order to reproduce these calculations, it is helpful to note
\begin{equation}
    \mathrm{Tr} \left( \mathcal{O} \psi \bar{\psi}\right)= - \bar{\psi} \mathcal{O} \psi = -i \psi^{\dagger} \gamma^0 \mathcal{O} \psi = - \psi^{\dagger} \sigma^z \mathcal{O} \psi,
\end{equation}
which readily implies
\begin{subequations}
\begin{align}
\mathrm{Tr} \left(\psi \bar{\psi}\right) &= -\bar{\psi}\psi = - \psi^{\dagger} \sigma^z  \psi \nonumber = 
\left(\begin{array}{cc} \psi^\dagger_0 & \psi^{\dagger}_1 \end{array} \right) 
\left( \begin{array}{cc} -1 & 0 \\ 0 & 1 \end{array} \right)  
\left( \begin{array}{c} \psi_0 \\ \psi_1 \end{array} \right)
\nonumber \\ &= - \psi^\dagger_0 \psi_0 + \psi^{\dagger}_1 \psi_1, \\ 
\mathrm{Tr} \left(\gamma^5 \psi  \bar{\psi}\right) &= -\bar{\psi} \gamma^5 \psi = - \psi^{\dagger} \sigma^z  \gamma^5 \psi = \psi^{\dagger} (-\sigma^y) \psi = 
\left(\begin{array}{cc} \psi^\dagger_0 & \psi^{\dagger}_1 \end{array} \right) 
\left( \begin{array}{cc} 0 & i \\ -i & 0 \end{array} \right)  
\left( \begin{array}{c} \psi_0 \\ \psi_1 \end{array} \right)
\nonumber \\ &= i \psi^\dagger_0 \psi_1 - i\psi^{\dagger}_1 \psi_0, \\
\mathrm{Tr} \left(\gamma_0 \psi  \bar{\psi}\right) &= -\bar{\psi} \gamma_0 \psi =- \psi^{\dagger} \sigma^z  \gamma_0 \psi =  - \psi^{\dagger} (-iI) \psi = 
\left(\begin{array}{cc} \psi^\dagger_0 & \psi^{\dagger}_1 \end{array} \right) 
\left( \begin{array}{cc} -i & 0 \\ 0 & -i \end{array} \right)  
\left( \begin{array}{c} \psi_0 \\ \psi_1 \end{array} \right)
\nonumber \\ &= -i \psi^\dagger_0 \psi_0 - i\psi^{\dagger}_1 \psi_1, \\
\mathrm{Tr} \left(\gamma_1 \psi  \bar{\psi}\right) &= -\bar{\psi} \gamma_1 \psi =- \psi^{\dagger} \sigma^z  \gamma_1 \psi = - \psi^{\dagger} (i \sigma^x) \psi = 
\left(\begin{array}{cc} \psi^\dagger_0 & \psi^{\dagger}_1 \end{array} \right) 
\left( \begin{array}{cc} 0 & i \\ i & 0 \end{array} \right)  
\left( \begin{array}{c} \psi_0 \\ \psi_1 \end{array} \right)
\nonumber \\ &= i \psi^\dagger_0 \psi_1 + i\psi^{\dagger}_1 \psi_0.
\end{align}
\end{subequations}
The condensate, pseudo-scalar, and current operators can be similarly evaluated and read
\begin{subequations}
\label{eq:UnsplitOps2}
\begin{align}
 C_I(t, x)  &\equiv m \bar{\psi}(t, x) \psi(t, x) = m\left( \psi^\dagger_0(t,x)\psi_0(t,x) -\psi^{\dagger}_1(t,x)\psi_1(t,x) \right)
 ,\\
 C_5(t, x)  &\equiv \bar{\psi}(t, x) \gamma^5 \psi(t, x) =  -i \psi^\dagger_0(t,x) \psi_1(t,x) + i\psi^{\dagger}_1(t,x) \psi_0(t,x),\\
 j_0(t, x) &\equiv \bar{\psi}(t, x) \tilde{\gamma}_0(t, x) \psi(t, x) = i\Omega(t,x) \left(  \psi^\dagger_0(t,x)\psi_0(t,x) +\psi^{\dagger}_1(t,x)\psi_1(t,x)\right),\\
 j_1(t, x) &\equiv \bar{\psi}(t, x) \tilde{\gamma}_1(t, x) \psi(t, x) =  -i\Omega(t,x) \left( \psi^\dagger_0(t,x)\psi_1(t,x) +\psi^{\dagger}_1(t,x)\psi_0(t,x)\right). 
\end{align}
\end{subequations}

\section{Minkowski-Spacetime Calculation of Hadamard-Renormalized Condensate}
\label{CondensateAppendix}
In this Appendix we calculate the Hadamard renormalized condensate $\langle C^\mathrm{ren}_I \rangle$ defined by \eqref{eq:QuadraticRenormalized}, in the inertial
vacuum of Minkowski spacetime. The equal-time Hadamard function can be found in e.g. \cite{PeskinSchroeder},
\begin{equation}
    G^{(1)}(t, x, x') = -\left(i \gamma^\mu \partial_\mu + m \mathbbm{1}\right) \int \frac{dp}{2\pi} \frac{1}{2\sqrt{p^2 + m^2}} e^{-i p r},
\end{equation}
where $r = x' - x$. Suppressing the artificial $t$ dependence, the point-split condensate defined by \eqref{eq:CurrentPointSplitCurved} is
\begin{align}
\mathbbm{C}_I(x, x') &= m \mathrm{Tr} \left(i \gamma^\mu \partial_\mu + m \mathbbm{1}\right) \int \frac{dp}{2\pi} \frac{1}{2\sqrt{p^2 + m^2}} e^{-i p r} \\
&= \int \frac{dp}{2\pi} \frac{m^2}{\sqrt{p^2 + m^2}} e^{-ipr}.
\end{align}
Now the modified Bessel function of the first kind, $K_0(\omega)$, may be expressed as
\begin{equation}
        K_0(\omega) = \frac{1}{2} \int db \frac{e^{i \omega b}}{\sqrt{b^2 + 1}},
\end{equation}
so that
\begin{equation}
\mathbbm{C}_I(x, x') = -\frac{m^2}{\pi} K_0(mr) = \frac{m^2}{2\pi} \ln \mu_0 \frac{1}{2} r^2 + \mathcal{O}(r^2)
\end{equation}
where $\mu_0 = \frac{1}{2}m^2 e^{2 \gamma_E}$ as in \eqref{eq:mu0}, and $\gamma_E$ is the Euler-Mascheroni constant. In the Minkowski metric, the locally determined term in \eqref{eq:HadCoord} is 
\begin{equation}
\mathbbm{C}^\mathrm{loc}_I(x, x') =  \frac{m^2}{2\pi} \ln \mu \frac{1}{2} r^2,
\end{equation}
and so choosing $\mu = \mu_0$ we have
\begin{equation}
 \langle C_I^\mathrm{ren} \rangle = \lim_{r \to 0} \left( \mathbbm{C}_I(x, x') - \mathbbm{C}^\mathrm{loc}_I(x, x') \right) = 0. 
\end{equation}

\bibliographystyle{plainnat}
\bibliography{FreePaper}
\end{document}